\documentclass[%
 reprint,
 amsmath,amssymb,
 aps,
prfluids,
onecolumn,
]{revtex4-2}

\usepackage{graphicx}
\usepackage{dcolumn}
\usepackage{bm}
\usepackage[caption=false,listofformat=subsimple,labelformat=simple]{subfig}

\usepackage{hyperref}

\begin{document}

\preprint{APS/123-QED}

\title{Impact-induced hardening in dense frictional suspensions}

\author{Pradipto}
\textbf{ \email{pradipto@yukawa.kyoto-u.ac.jp}}
\author{Hisao Hayakawa}%

\affiliation{%
 Yukawa Institute for Theoretical Physics, Kyoto University, Kitashirakawaoiwake-cho, Sakyo-ku, Kyoto 606-8502, Japan
}%




\date{\today}

\begin{abstract}
We numerically study the impact-induced hardening in dense suspensions.
We employ the lattice Boltzmann method and perform simulations of dense suspensions under impacts, which incorporate the contact between suspended particles with the free surface of the suspension.
Our simulation for a free-falling impactor on a dense suspension reproduces experimental results, where rebound takes place for frictional particles at high-speed impact and high volume fraction shortly after the impact before subsequently sinking.
We found that the shear stress of the suspension is not affected by the impact, which clearly distinguishes the impact-induced hardening from the discontinuous shear thickening.
Instead, we found the existence of a localized region with distinctively high value of normal stress corresponding to the dynamically jammed region.
Our simulation indicates that the frictional interaction between suspended particles is important for the impact-induced hardening to maintain the dynamically jammed region.
Furthermore, persistent homology analysis successfully elucidates the topological structure of force chains.
\end{abstract}

\maketitle
\section{Introduction} 

A dense suspension can behave as a fluid or a solid depending on the situation.
One of the examples of this non-Newtonian behavior is a running person can stay afloat on top of the suspensions, while a walking person sinks.
The phenomenon that the suspension exhibits solid-like response under fast impact also has practical applications, such as protective vests when it is combined with fibers as a composite material \cite{lee2003,nam2005}.
Some efforts have been made to reproduce such hardening during impact through experiments.
Waitukaitis and Jaeger discovered a dynamically jammed region which is like a solid plug beneath the impactor \cite{waitukaitis2012}. 
They proposed an added mass effect to explain the solidification induced by the impact.
A similar hardening process is also observable in fractures on a thin layer of suspension under an impact \cite{roche2013}.
Then, by using a high-speed ultrasound imaging, Han et al. measured the sound speed and visualized the flow field of the suspension \cite{han2015}.
Their measurement of the sound speed showed no increase in local volume fraction.
In addition, they suggested that the mechanism behind such solidification in dense suspensions under impact is related to the jamming by shear, instead of densification.
Moreover, a series of constant speed penetration experiments indicated that the solidification occurs when the dynamically jammed region span from the impactor to the boundaries \cite{maharjan2018,allen2018,mukhopadhyay2018}.
Finally, the impact-induced hardening can also be characterized by dropping an impactor into a dense suspension \cite{egawa2019}.

Some people use the discontinuous shear thickening (DST) to explain the impact-induced hardening \cite{lee2003,allen2018,mukhopadhyay2018}, while the connection between these two processes is unclear \cite{brown2014}.
Actually, there are some differences between these two processes.
First, DST is observed in the dense suspensions undergoing steady shear, while the impact-induced hardening is a transient process undergoing normal compression.
Second, the flow field of dense suspensions under impact is inhomogeneous \cite{han2015}.
This is in contrary to the common DST which is anisotropic but still homogeneous \cite{pradipto2020}.
Therefore, to make a further distinction between the impact-induced hardening and the DST, we need a detailed study of impact-induced hardening.
Some papers also suggested some similarities between the shear jamming and the impact-induced hardening \cite{han2015,peters2016}, but the connection between two processes is also unclear.
To clarify the relation is also one of the purpose of this study.

Even though the aforementioned experiments have already visualized the displacement and flow fields \cite{waitukaitis2012, han2015}, and measured the stress exerted on the impactor \cite{maharjan2018}, any experimental measurement on the shear and normal stresses fields of dense suspensions under impact has not been reported yet.
On the other hand, the local distribution of the stress can be calculated and visualized through particle-based suspensions simulations \cite{seto2013,mari2014,pradipto2020}.
Moreover, numerical simulation is an important tool to understand the microscopic mechanism behind exotic phenomena in suspensions since the motion of the suspended particles is not visible in three-dimensional experiments, unlike in 2-dimensional dry granular materials where the force acting on each grain can be visualized with the photoelastic disks \cite{clark2015}.
However, a particle-based simulation of a free-falling impactor hitting a suspension has not been reported so far because of the difficulty of simulating suspension with free surface.
As far as we know, the first fluid-based simulation of suspensions under impact has been conducted recently in Ref. \cite{baumgarten2019}, where the authors successfully reproduced various interesting processes for suspensions under impact, such as the viscoelastic response of a dense suspension to a rotating wheel.
Since, however, their fluid simulations with a constitutive equation with some fitting parameters cannot capture the particle dynamics, the mechanism behind impact-induced hardening on the microscopic level remains elusive.
In this paper, we utilize the lattice Boltzmann method (LBM) to perform a simulation to capture particle dynamics \cite{ladd1994a,ladd1994b,succi2001} .
By using LBM for hydrodynamic calculation, combined with the particle interaction scheme in Refs. \cite{seto2013,mari2014}, we have already analyzed the shear jamming and DST on dense suspensions under simple and oscillatory shear \cite{pradipto2020}.
Now, we upgrade the simulation scheme by incorporating the free surface of the suspensions as first demonstrated in Refs. \cite{svec2012,leonardi2014,leonardi2015}. 
Then, we can characterize the impact-induced hardening by capturing the particle dynamics of a suspension.

The outline of this paper is as follows.
In Sec. II, we explain our simulation method briefly.
In Sec. III we present the simulation results of free-falling impactor onto dense suspensions which include the kinematics of the impactor, phenomenological model, and visualization of local quantities such as the stress tensor, local volume fraction, and displacement.
We also present the persistent homology analysis to capture the topological structure of force chains.
In Sec. IV, we summarize our results and discuss future perspectives. 
In Appendix A, we describe the details of the LBM involving suspensions with free surface \cite{svec2012,leonardi2014,leonardi2015}.
In Appendix B, we discuss the system size dependence of our simulations.
In Appendix C, we illustrate how to implement the persistent homology analysis for force networks observed in our simulation.

\section{Simulation method}
We employ the LBM involving suspensions with free surface of liquid.
The details of our method are explained in Appendix A.
Due to the discrete nature of the LBM, we take the lattice unit $\Delta x = 0.2 a_{\text{min}}$ ($a_{\text{min}}$ is the radius of the smallest particle) for the calculation of hydrodynamic fields. 
The suspended particles in LBM are represented as a group of solid nodes, while the surrounding fluids are represented by fluid nodes.
The hydrodynamic field is calculated from the time evolution of the discrete distribution function at each fluid node.
The choice of a smaller lattice unit means a smoother surface of the sphere, but it becomes expensive.
We select the lattice unit $\Delta x = 0.2 a_{\text{min}}$, where it still give sufficient accuracy but still not computationally expensive as shown in the previous LBM for suspensions literatures \cite{ladd1994a,ladd1994b,nguyen2002}.
Then, to simulate the free surface of the fluid, it is necessary to introduce interface nodes between the fluid and gas nodes, as explained in Appendix A \cite{svec2012,leonardi2014,leonardi2015}.

\subsection{Discrete element method for suspended particles}

Equations of motion and the torque balance of particle $i$ are, respectively, given by 
\begin{equation}
	m_i \frac{d \bm{u}_i}{dt} = \bm{F}_i^c + \bm{F}_i^h + \bm{F}_i^{\rm lub} + \bm{F}_i^r + \bm{F}_i^g,
	\label{eom_part}
	\end{equation}
\begin{equation}
	I_i \frac{d \bm{\omega}_i}{dt} = \bm{T}_i^c +  \bm{T}_i^{\rm lub} + \bm{T}_i^{h}.
	\label{eot_part}
	\end{equation}
Here, $\bm{u}_i$, $\bm{\omega}_i$, $m_i$, and $I_i=(2/5) m_i a_i^2$ (with $a_i$ the radius of particle $i$), are the translational velocity, angular velocity, mass, and the moment of inertia of particle $i$, respectively.
$\bm{F}_i^g = - m_i g \hat{\bm{z}}$ is the gravitational force acting on the suspended particle $i$, where $g$ is the gravitational acceleration and $\hat{\bm{z}}$ is the unit vector in the vertical direction. 
Note that our LBM accounts for both the short lubrication force $\bm{F}_i^{\rm lub}$ and torque $\bm{T}_i^{\rm lub}$, as well as the long-range parts of the hydrodynamic force $\bm{F}_i^h$ and torque $\bm{T}_i^h$ \cite{nguyen2002,pradipto2020}.
The long-range parts ($\bm{F}_i^h$ and $\bm{T}_i^h$) are calculated using the direct forcing method (See Appendix A), while the lubrication force $\bm{F}_i^{\rm lub}$ and torque $\bm{T}_i^{\rm lub}$ are expressed by pairwise interactions as $\bm{F}_{i}^{\rm lub} = \sum_{j \neq i} \bm{F}_{ij}^{\rm lub}$ and $\bm{T}_{i}^{\rm c} = \sum_{j \neq i} \bm{T}_{ij}^{\rm lub}$, respectively \cite{seto2013,mari2014,nguyen2002,pradipto2020}.
The contact force $\bm{F}_i^c$ and torque $\bm{T}_i^c$ of particle $i$ are also expressed by pairwise interaction as $\bm{F}_{i}^{\rm c} = \sum_{j \neq i} \bm{F}_{ij}^c$ and $\bm{T}_{i}^{\rm c} = \sum_{j \neq i} \bm{T}_{ij}^c$ and are computed using the linear-dashpot model with Coulomb friction rules and friction coefficient $\mu$ \cite{luding2008}.
Finally, $\bm{F}_i^r$ is the electrostatic repulsive force, also expressed by pairwise interactions as $\bm{F}_{i}^{\rm r} = \sum_{j \neq i} \bm{F}_{ij}^r$.
This force arises from the double layer to prevent particles from clustering with neglecting the Brownian force\cite{pradipto2020,mari2014}.
The explicit expressions of $\bm{F}_{ij}^{lub}$,$\bm{T}_{ij}^{lub}$, $\bm{F}_{ij}^{c}$, $\bm{T}_{ij}^{c}$, and $\bm{F}_{ij}^{r}$ can be found in Ref. \cite{pradipto2020}.
Throughout this paper, we have adopted the perfect density matching between the solvent and suspended particles, where the densities of solvent and particles satisfy the relation $\rho_f=\rho_p$ with the densities of a suspended particle $\rho_p$ and solvent fluid $\rho_f$.

The impactor is a solid spherical object with density $\rho_I = 4 \rho_f$.
The force and torque acting on the impactor are, respectively, given by 
\begin{equation}
\bm{F}^{I} = \bm{F}^{I,h} + \bm{F}^{I,\rm lub} + \bm{F}^{I,c} + \bm{F}^{I,g},
	\end{equation}
\begin{equation}	
\bm{T}^{I} = \bm{T}^{I,h} +  \bm{T}^{I,c} +  \bm{T}^{I,\rm lub}.
	\end{equation}
$\bm{F}^{I,g} = - m_I g \bm{\hat{z}}$ is the gravitational force acting on the impactor with mass $m_I$.
The contact force $\bm{F}^{I,c}$ and torque $\bm{T}^{I,c}$ arise from the interactions with the suspended particles, also calculated with the linear-dashpot model with Coulomb friction rules.
The hydrodynamic force $\bm{F}^{I,h}$ and torque $\bm{T}^{I,h}$ are calculated using the bounce-back rule that satisfies the no-slip boundary condition between the fluid and the surface of the impactor \cite{ladd1994a,ladd1994b}.
Here, the bounce-back rule is implemented by treating the surface of the impactor as boundary nodes.
When the LBM discrete distribution function streams from fluid nodes to the boundary nodes, it is reflected, which we call it the bounce-back rule.
Then, the hydrodynamic force on each node is calculated from the momentum transferred in that bounce-back process.
We cannot use the direct forcing method in Eqs. \eqref{eom_part} and \eqref{eot_part} as explained in Appendix A for the impactor since the method requires the whole surface of the impactor to be surrounded by liquid.
The lubrication force $\bm{F}^{I,\rm lub}$ and torque $\bm{T}^{I,\rm lub}$ are calculated with a similar manner as suspended particles.

\subsection{Simulation setup}
\begin{figure*}[htbp]
	\centering
	\subfloat[]{\label{fig:1a}%
	  \includegraphics[width=0.25\textwidth]{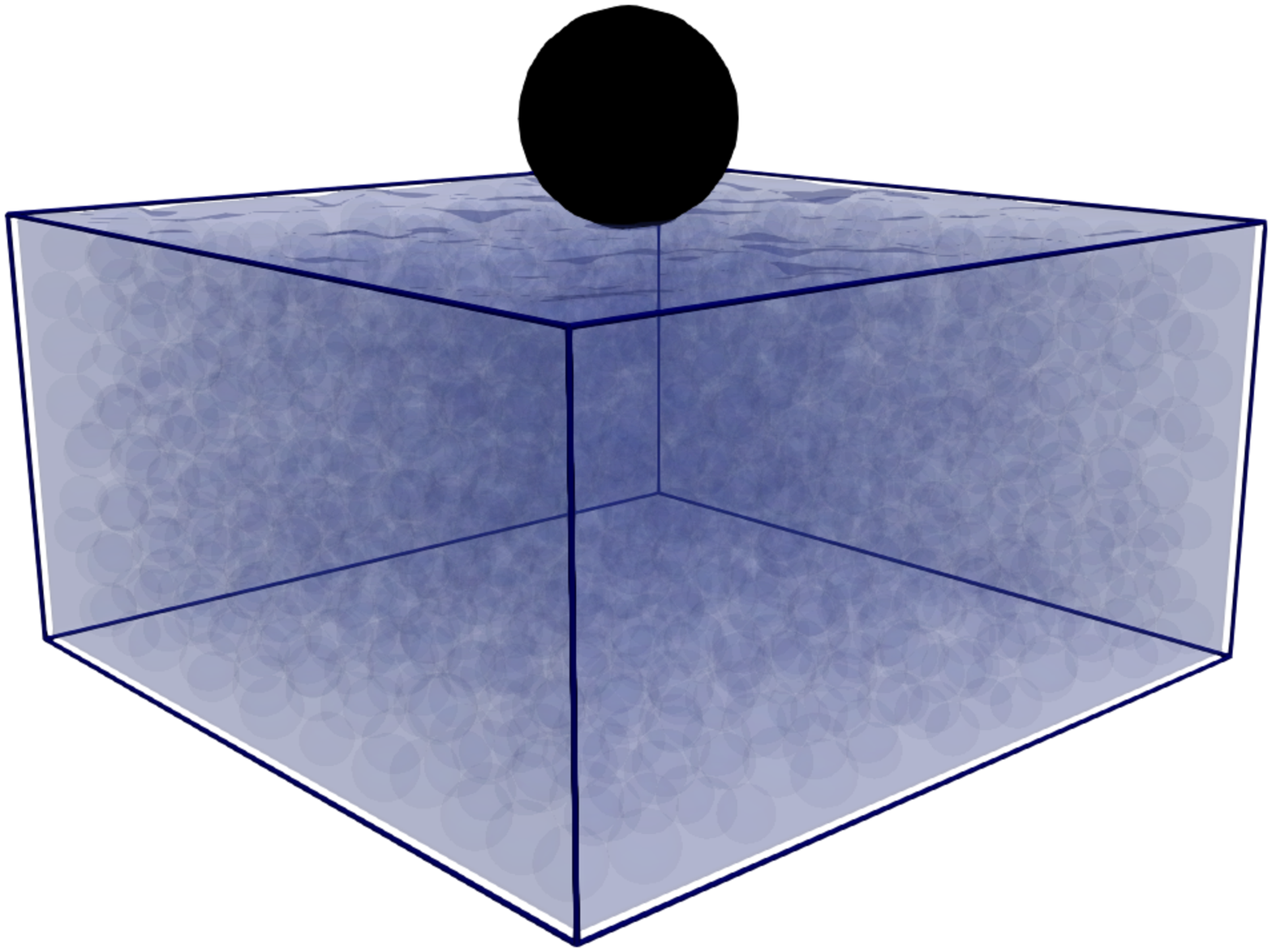}
	}
	\subfloat[]{\label{fig:1b}%
	  \includegraphics[width=0.25\textwidth]{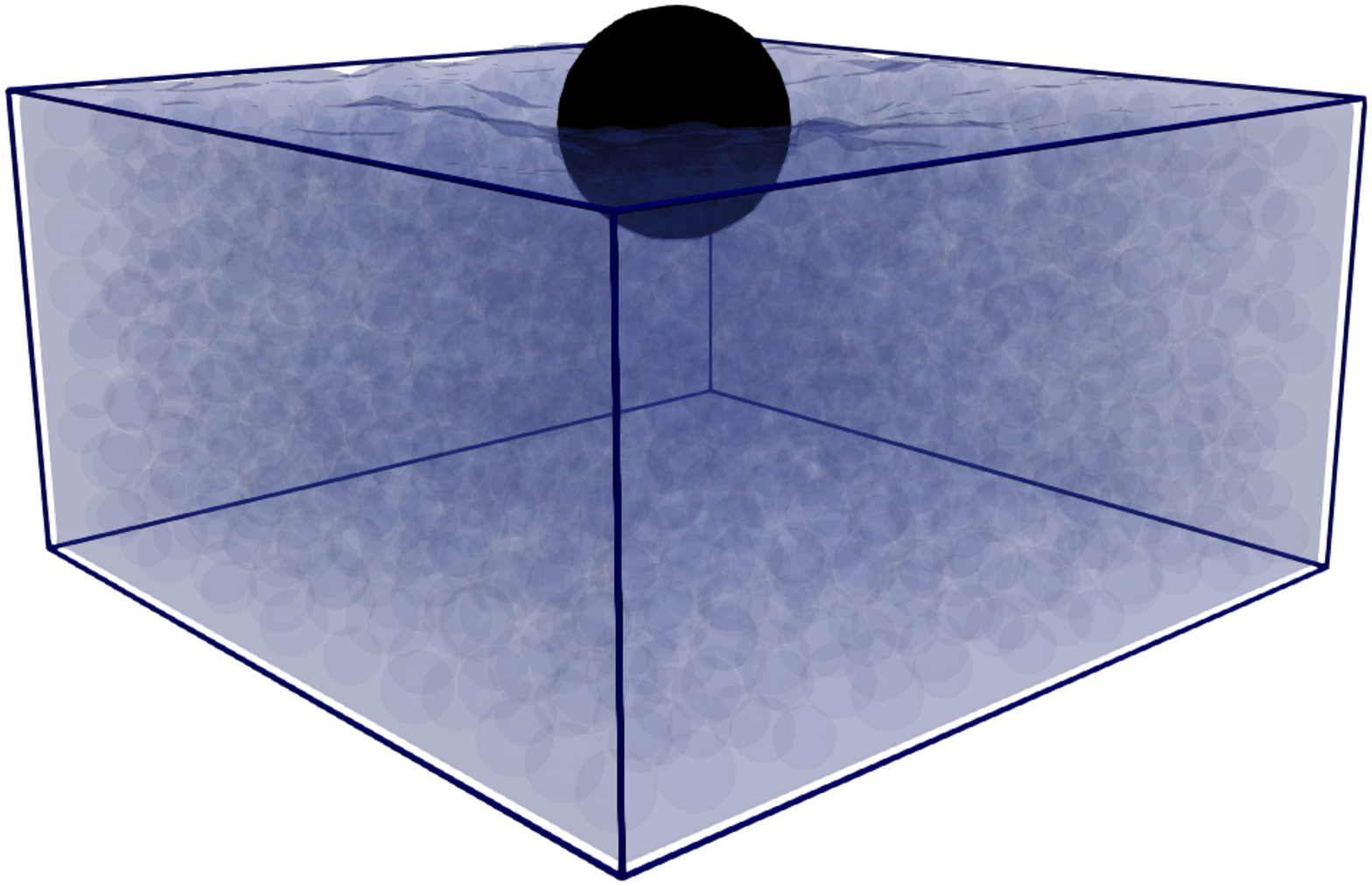}
	}
	\subfloat[]{\label{fig:1c}%
	  \includegraphics[width=0.25\textwidth]{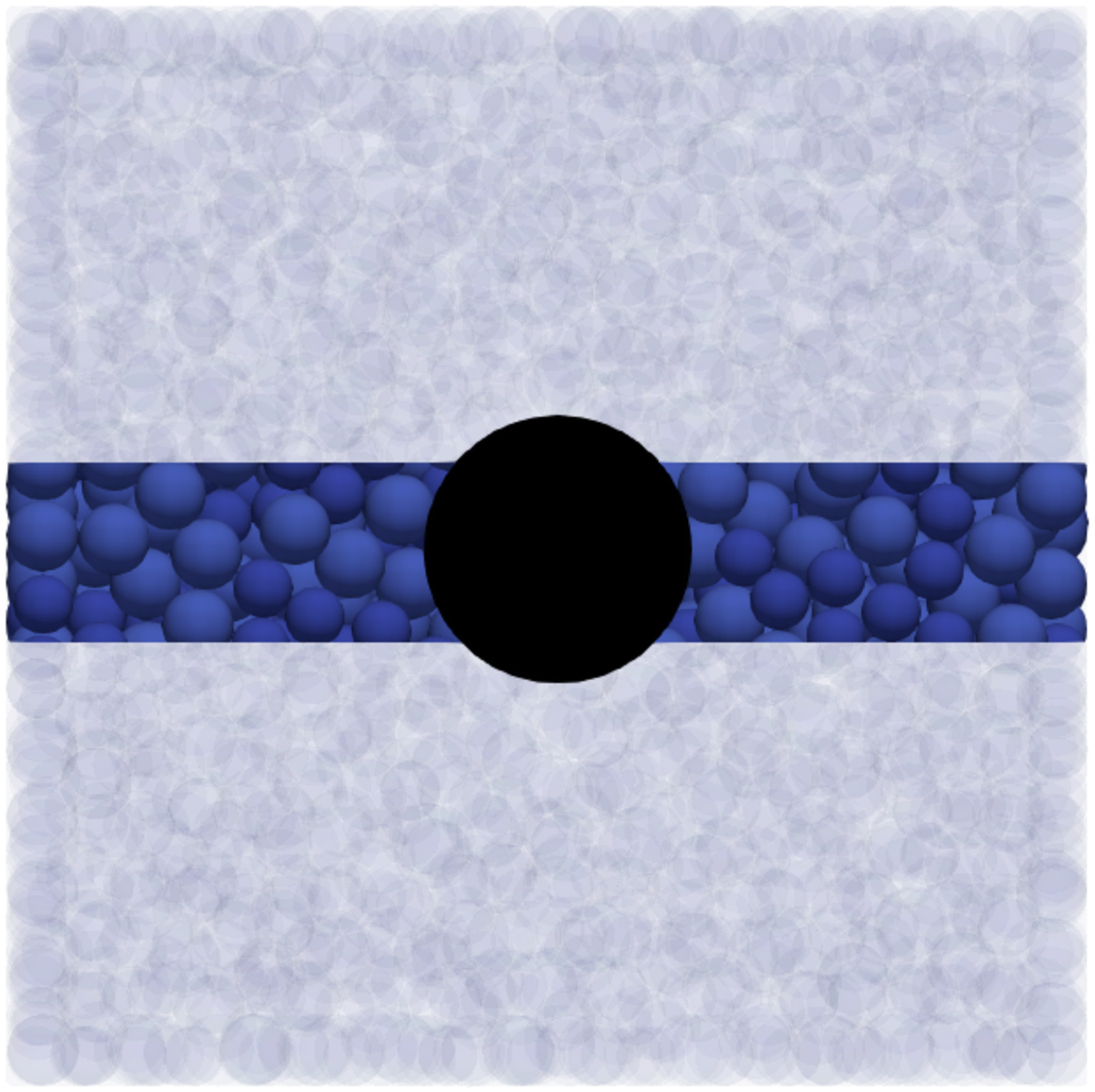}
	}
	
	\subfloat[]{\label{fig:1d}
	\includegraphics[width=0.9\textwidth]{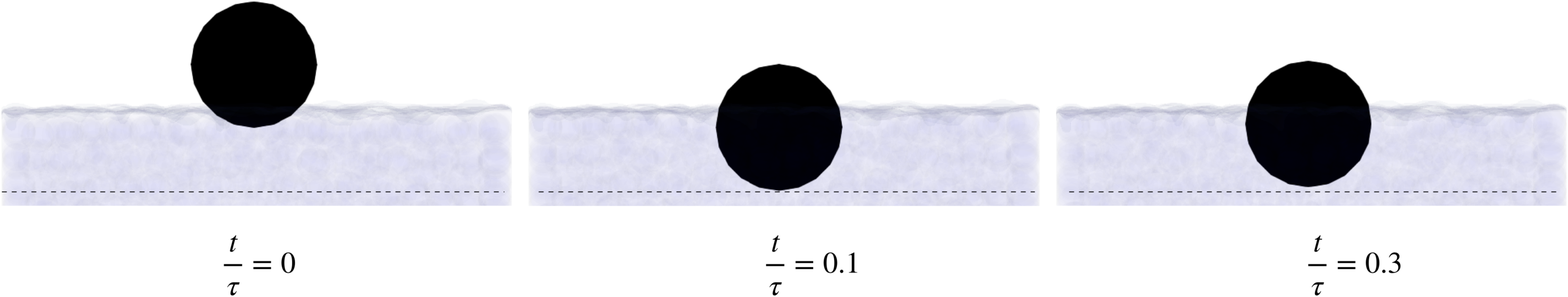}
	}
	\caption{(a) A snapshot of our 3D simulation for $\phi = 0.54$, $\mu=1$, and $u_0/u^{*} = 4.26$ at $t/\tau = 0$. (b) The time evolution of (a) at $t/\tau = 0.1$. (c) A top view of the sliced region. (d) Successive snapshots of the impactor in a quasi-two-dimensional slice of container as in (c), where the dashed lines mark the maximum penetration.}
	\end{figure*}

Suspended particles (with bidispersity ratio $a_{\text{max}} = 1.2 a_{\text{min}}$, where the radii of the large and small particles are $a_{\text{max}}$ and $a_{\text{min}}$, respectively) are confined into a rectangular box ($W \times D \times H$) with smooth walls.
The system contains $N=2000$ particles for most simulations and sometimes contains N=1200. 
The radius of the impactor $a_I$ is $a_I=4.5 a_{\rm min}$ for $N=2000$ and $a_I = 3.75 a_{\rm min}$ for $N=1200$.
We fix $W/a_I = 8$, $D/a_I = 8$, and $H/a_I = 4$ except for the argument in Sec. III.B.
The system size dependence is discussed in Appendix B.
The impactor is released from various heights $H_0$ that correspondsimpactor to the impact velocity as $u^{I}_{0} = \sqrt{2 g H_0}$, which also specifies the units of time in our simulation $\tau = \sqrt{a_{\text{min}}/2g}$, units velocity $u^{*} = \sqrt{2 g a_{\text{min}}}$, units of force $F_0=\frac{4}{3} \pi \rho_f a_{\rm min}^3 g$, and units of stress $\sigma_0 = F_0/a_{\rm min}^2$.

\section{Results}

\subsection{Impact-induced hardening}

\begin{figure*}[htbp]
\subfloat[]{\label{fig:2a}%
  \includegraphics[width=0.36\textwidth]{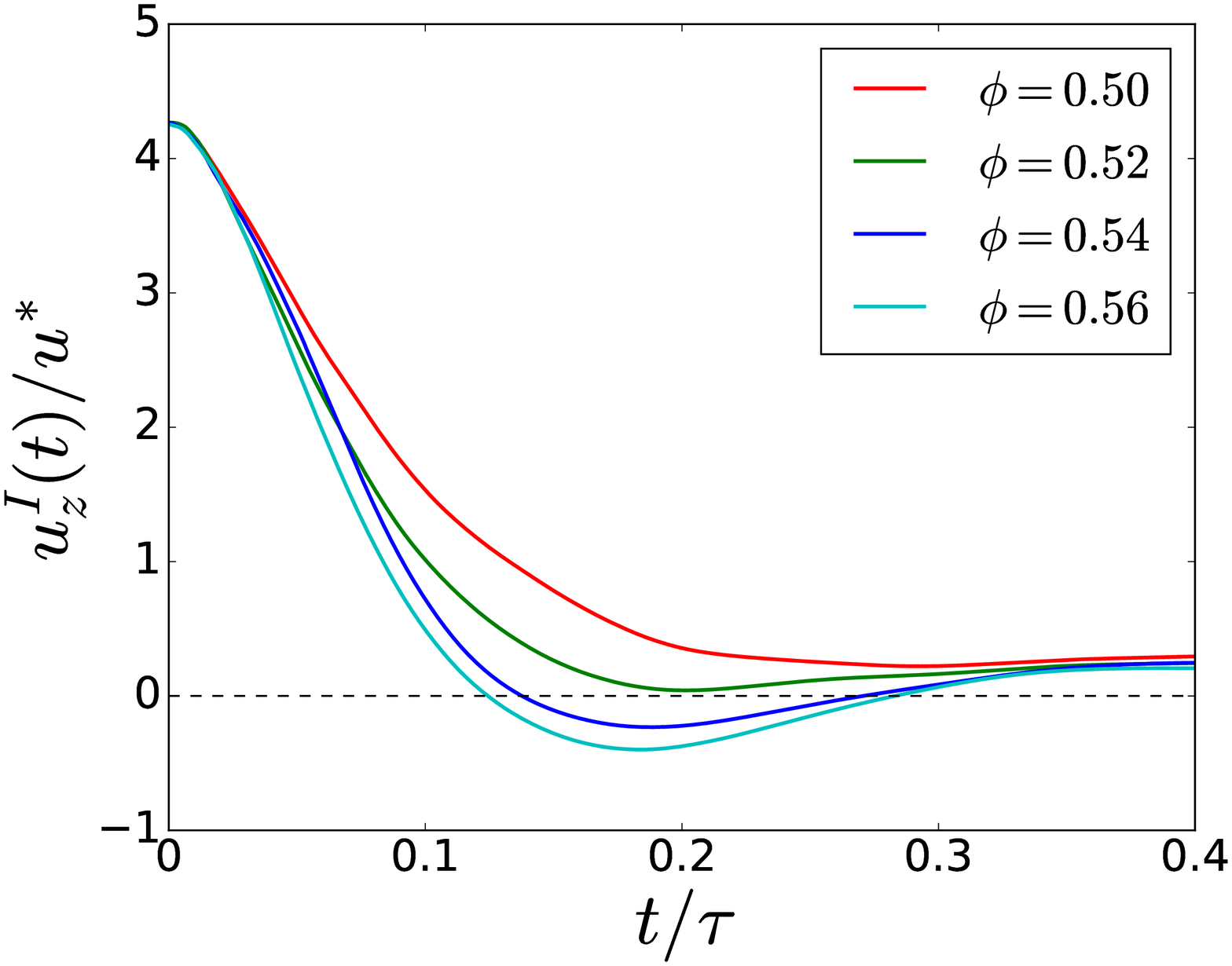}
}
\subfloat[]{\label{fig:2b}%
  \includegraphics[width=0.36\textwidth]{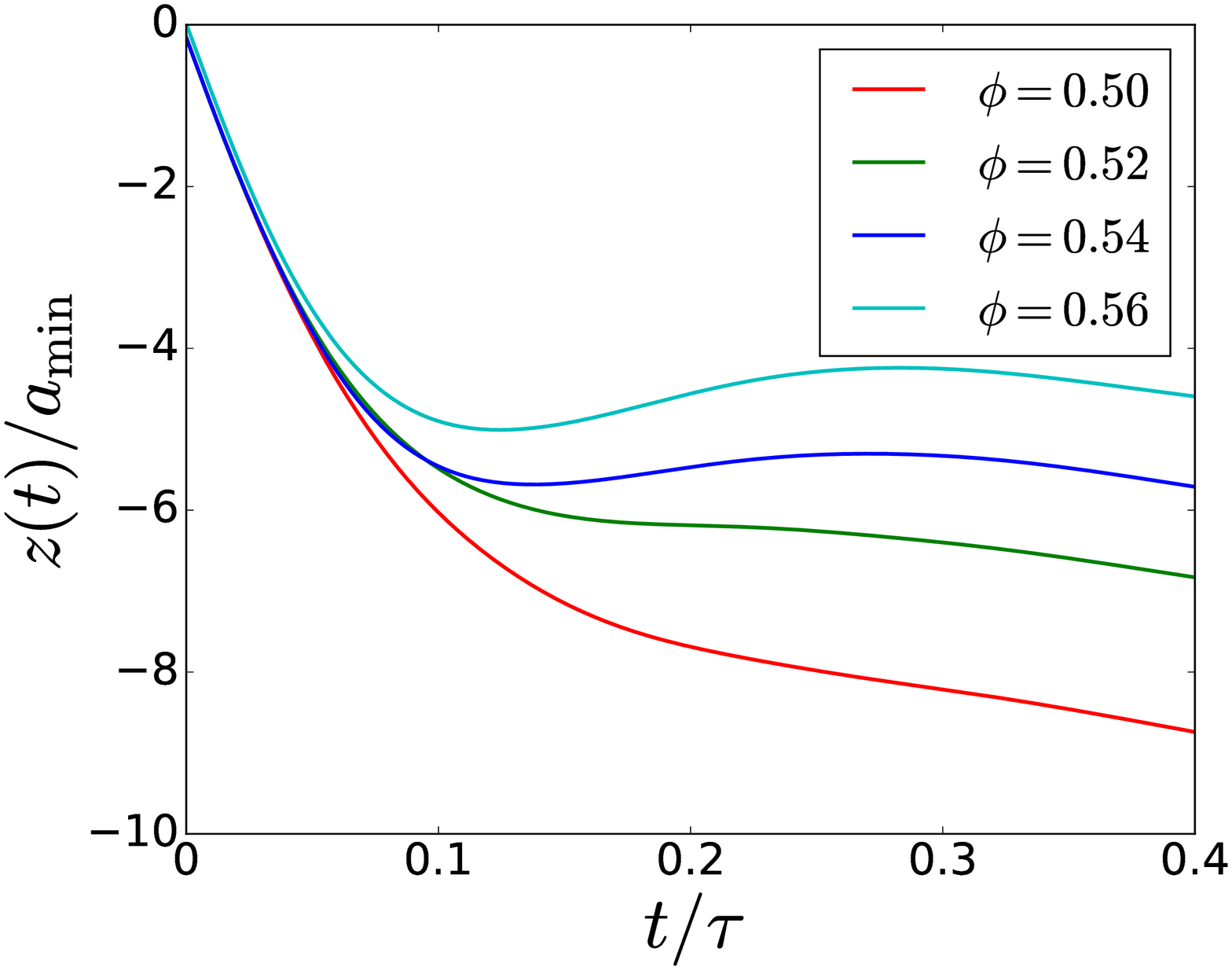}
}
\caption{(a) Plots of impactor speeds in the $z$-direction, $u^I_z (t)/u^{*}$, against time for various volume fractions $\phi$. (b) Plots of the heights of the impactor against time for various volume fractions $\phi$. Both results are obtained by using 2000 frictional particles whose friction constant is $\mu=1$.}
\end{figure*}

Three dimensional snapshots of a free-falling impactor simulation can be seen in Figs. \ref{fig:1a} and \ref{fig:1b}.
For visualization, we slice the system in the middle of the container, as shown in Fig. \ref{fig:1c}.
By looking at the successive motion of the impactor from Fig. \ref{fig:1d}, where we set time $t=0$ and height $z=0$ at the moment of impact, one can confirm that the impactor penetrates and slightly rebounds after the maximum penetration.
This rebound motion of the impactor can be clearly observed by watching the Supplemental movie \cite{supp_movie}.
We plot the impactor speeds $u^I_z (t)/u^{*}$ against time for various volume fractions of suspended particles $\phi$ in Fig. \ref{fig:2a}.
The vertical position of the impactor $z(t) / a_{\rm min}$ for various volume fractions $\phi$ can be seen in Fig. \ref{fig:2b}.
Both results are obtained by using 2000 frictional particles with $\mu=1.0$.
One can observe the rebound of the impactor ($u^I_z (t)/u^{*} < 0$) for $\phi \geq 0.54$, which agrees semi-quantitatively with the free-falling impactor experiment \cite{egawa2019}. 
This rebound is the instance of the impact-induced hardening of the suspension shortly after the impact.
After the rebound, the suspension relaxes, and the impactor starts to sinks.
Note that $\phi$ for rebounds might be a little higher than that in the experiment \cite{egawa2019}.

\begin{figure*}[htbp]
\subfloat[]{\label{fig:3a}%
  \includegraphics[width=0.36\textwidth]{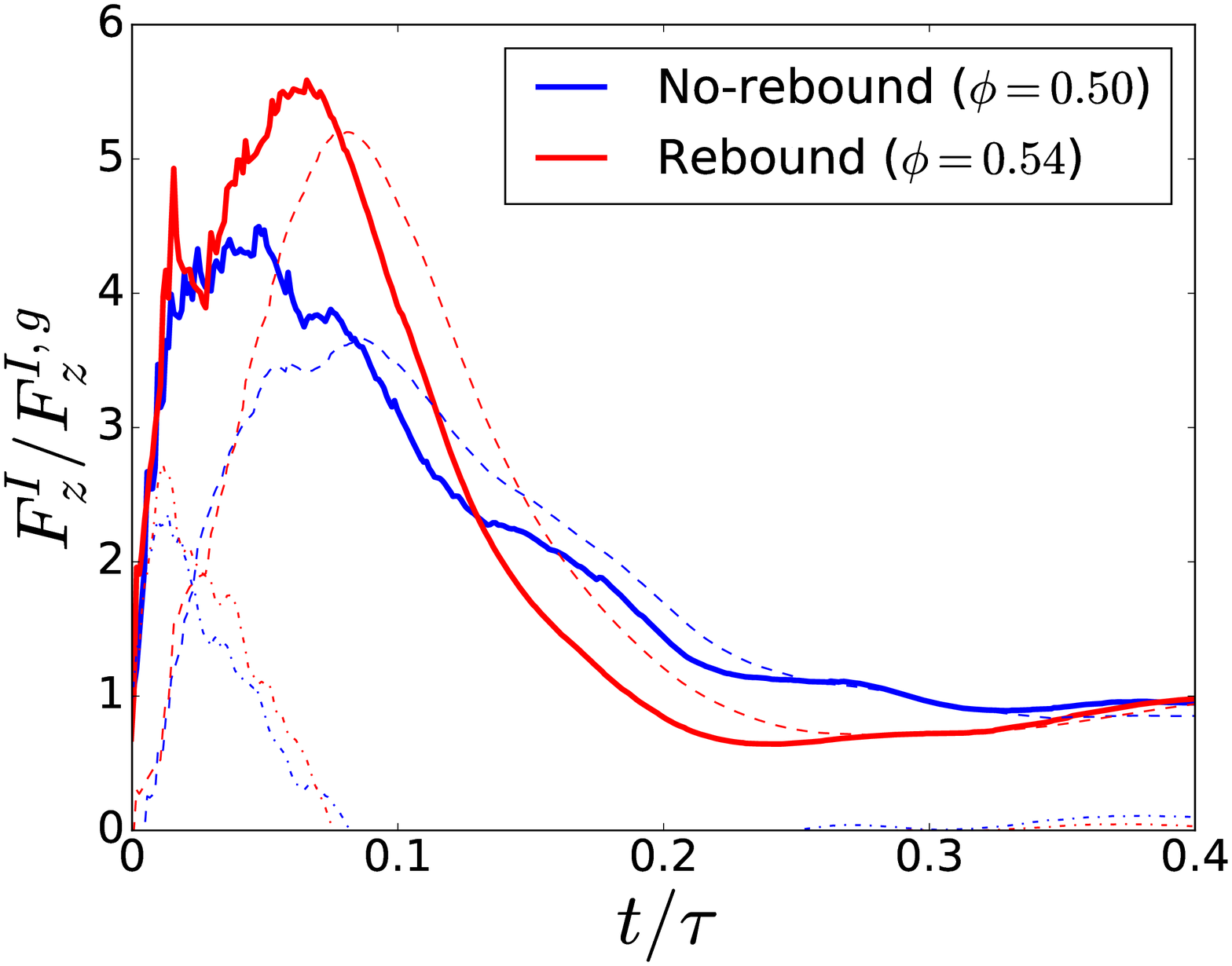}
}
\subfloat[]{\label{fig:3b}%
  \includegraphics[width=0.36\textwidth]{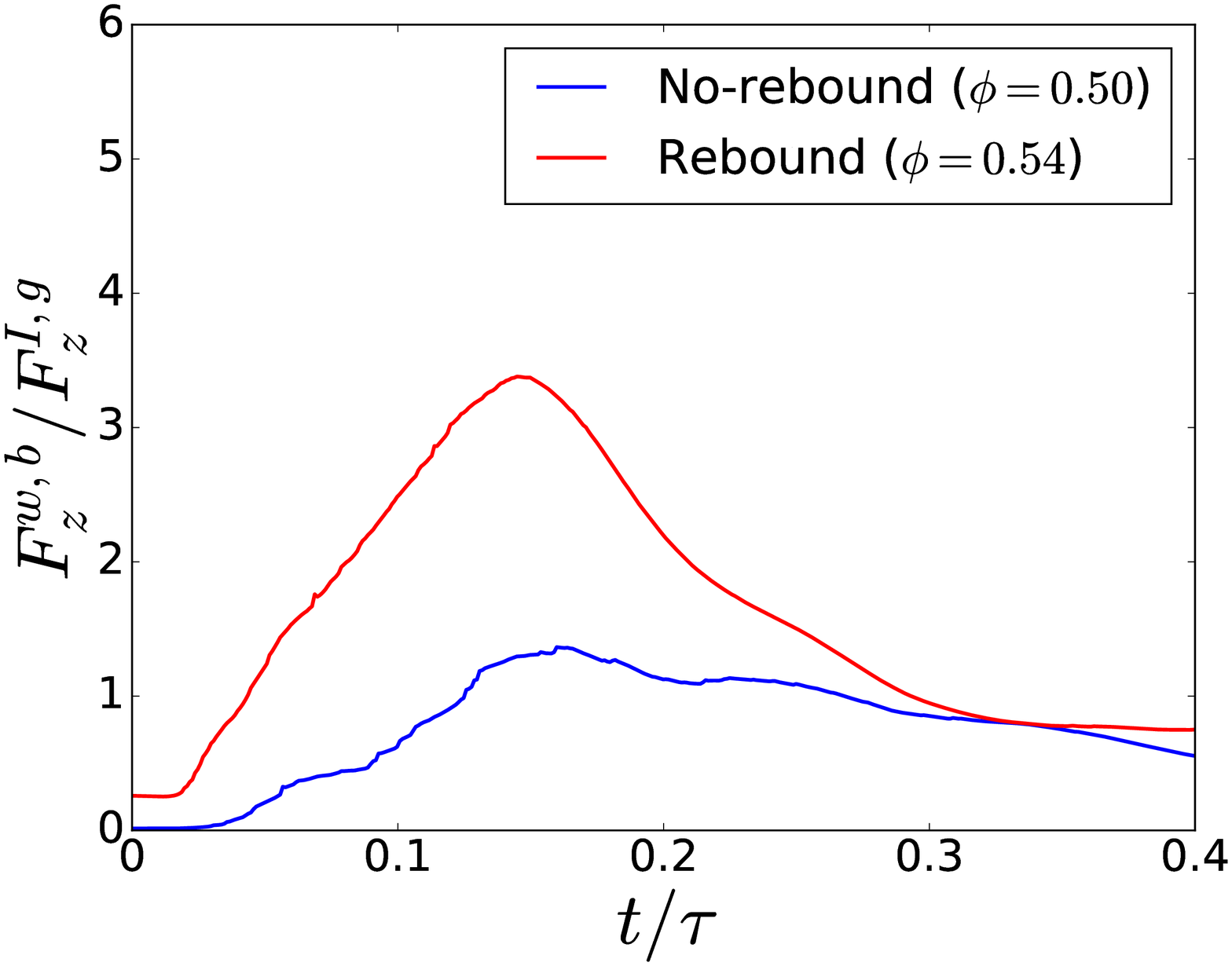}
}
\caption{Plots of the force for both rebound ($\phi = 0.54$) and no-rebound ($\phi = 0.50$) cases with $N=2000$. (a) Force exerted on the impactor, where the solid lines are the total force, dashed lines represent the contact contributions, dot-dashed lines represent the hydrodynamic contributions, and (b) the total force exerted on the bottom wall. All results are obtained for $\mu=1$ and $u_0/u^{*} = 4.26$.}
\end{figure*}

In Fig. \ref{fig:3a}, we plot the time evolution of total forces exerted on the impactors for both rebound and no-rebound cases for $\mu=1.0$ and $N=2000$.
One can see that the maximum exerted force for the rebound case is larger than that for the no-rebound case.
We find that the peak of the contact force is located slightly after the peak of the total force, which follows the weaker peak from the hydrodynamic contribution.
The time difference between these two peaks is not large so that they merge to a single peak in the total force.
In an experiment with rod impactor, two peaks in the acceleration of the impactor are observed for deep suspensions while for shallower suspensions, in which rebound takes place, the separation between peaks is not detectable \cite{waitukaitis2012}.
Thus, we confirm that the second peak in Ref. \cite{waitukaitis2012} is originated from the contact contribution. 
Moreover, they also observed the second peak when the impact force is transmitted to the boundary.
To clarify this, we plot the force exerted on the bottom wall in Fig. \ref{fig:3b}.
Compared to the force exerted on the impactor, one can see a clearer distinction between the rebound and no-rebound cases, where the force exerted on the bottom wall for rebound case is about three times larger than the no-rebound case.
This indicates that the hardening takes place when the contact force network percolates from the impactor to the boundaries, which is consistent with the picture in Ref. \cite{maharjan2018}.

\subsection{Hertzian contact model}

\begin{figure}[htbp]
	 \includegraphics[width=0.36\linewidth]{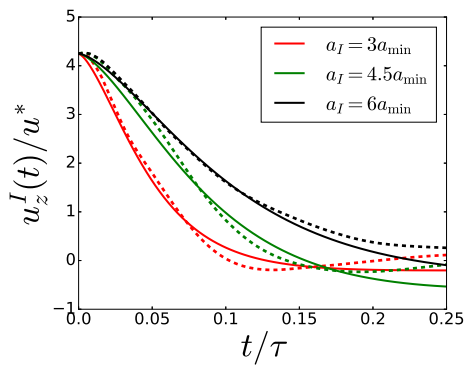}
	 \caption{Plots of the speeds of impactors in the $z$-direction $u^I_z (t)/u^{*}$ against time (dashed lines) and the solution of Eq. \eqref{eq:eom} (solid lines) for $\phi = 0.54$ and $N=2000$ with fitting parameters $A=1.64 \times 10^{5} m_0/(a_{\rm min} \tau^2)$ and $B= 6.48 \times 10^{4} m_0/(a_{\rm min} \tau)$ for various $a_I$ and  $m_0 = 4 \pi a_{\rm min}^3 \rho_f/3$.}
	 	 \label{fig:4} 
	\end{figure}
	
Let us explain the time evolution of the impactor velocity.
As can be seen in Fig. \ref{fig:3a}, the main contribution of the force acting on the impactor is the elastic contact force.
Also, the liquid surface behaves as an elastic sheet when the penetration of the impactor is small.
Therefore we may model of the impactor motion by using the Hertzian contact theory between the impactor and a sufficiently large elastic sheet supported by the side walls \cite{kuwabara1987,brilliantov1996}. 
To verify our picture, we vary the impactor radius $a_I$ and show how the impactor dynamics depends on its radius on Fig. \ref{fig:4}.
The equation motion for the deformation $h$ of the Hertzian contact is written as
\begin{equation}
	m_I \frac{d^2 h}{ dt^2} = - A \sqrt{a_I} h^{\frac{3}{2}} - B \sqrt{a_I} h^{\frac{1}{2}} \frac{d h}{d t}, \label{eq:eom}
	\end{equation}
where $A$ and $B$ are fitting parameters which correspond to the elastic modulus and viscosity, respectively.
In Fig. \ref{fig:4}, we plot the results of the simulation alongside with the solutions of Eq. \eqref{eq:eom}.
One can see that the results of the simulation agree with the model shortly after the impact.
We also clarified that the impact speed $u^I_z(t)$ clearly depends on the radius of impactor, which cannot be explained in the linear spring model used in Ref. \cite{egawa2019}.
One of the limitations of this model is that it completely ignores the hydrodynamic contributions.
Therefore, it deviates after the elastic response (rebound).
To describe sinking processes of the impactor, including the stop-go cycles of a sinking impactor \cite{vonkann2011}, we need a different approach \cite{maharjan2017}.
However, such behavior and analysis are beyond the scope of this paper.

\subsection{Phase diagram with $\mu=1$}

\begin{figure}[htbp]
	 \includegraphics[width=0.36\linewidth]{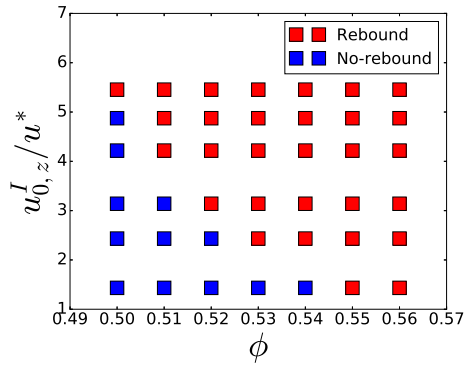}
	 \caption{Phase diagram showing whether the impactor rebounds before sinking as a function of the volume fraction $\phi$ and the impact speed $u^{I}_{0,z}$.}
	 \label{fig:5}
	\end{figure}

We observe that the impact-induced hardening depends on the impact speed $u^{I}_{0,z}$, as shown in the diagram of Fig. \ref{fig:5}, where we have fixed the friction coefficient as $\mu=1$.
Due to the limitation of our computational resources, the data for this phase diagram are obtained with $N=1200$ particles.
The finite-size effects is discussed in Appendix B.

Some previous papers reported that impact-induced hardening depends on impact speed \cite{maharjan2018,baumgarten2019}.
Note that the highest rebound volume fraction ($\phi=0.56$) in our simulation is still below the frictional ($\mu = 1$) jamming fraction $\phi^{\mu=1}_J \approx 0.585$ \cite{silbert2010}, whereas rebound takes place for $0.50 \leq \phi \leq 0.56$.
This range is similar to the observed volume fractions for the DST under simple shear in numerical simulations \cite{seto2013,mari2014,pradipto2020}.
However, one should recognize that two processes are different since impact-induced hardening is a heterogeneous and transient process, while shear thickening is a homogenous steady states process.

\subsection{Roles of the friction between particles}

\begin{figure}[htbp]
\subfloat[]{
	 \includegraphics[width=0.3\linewidth]{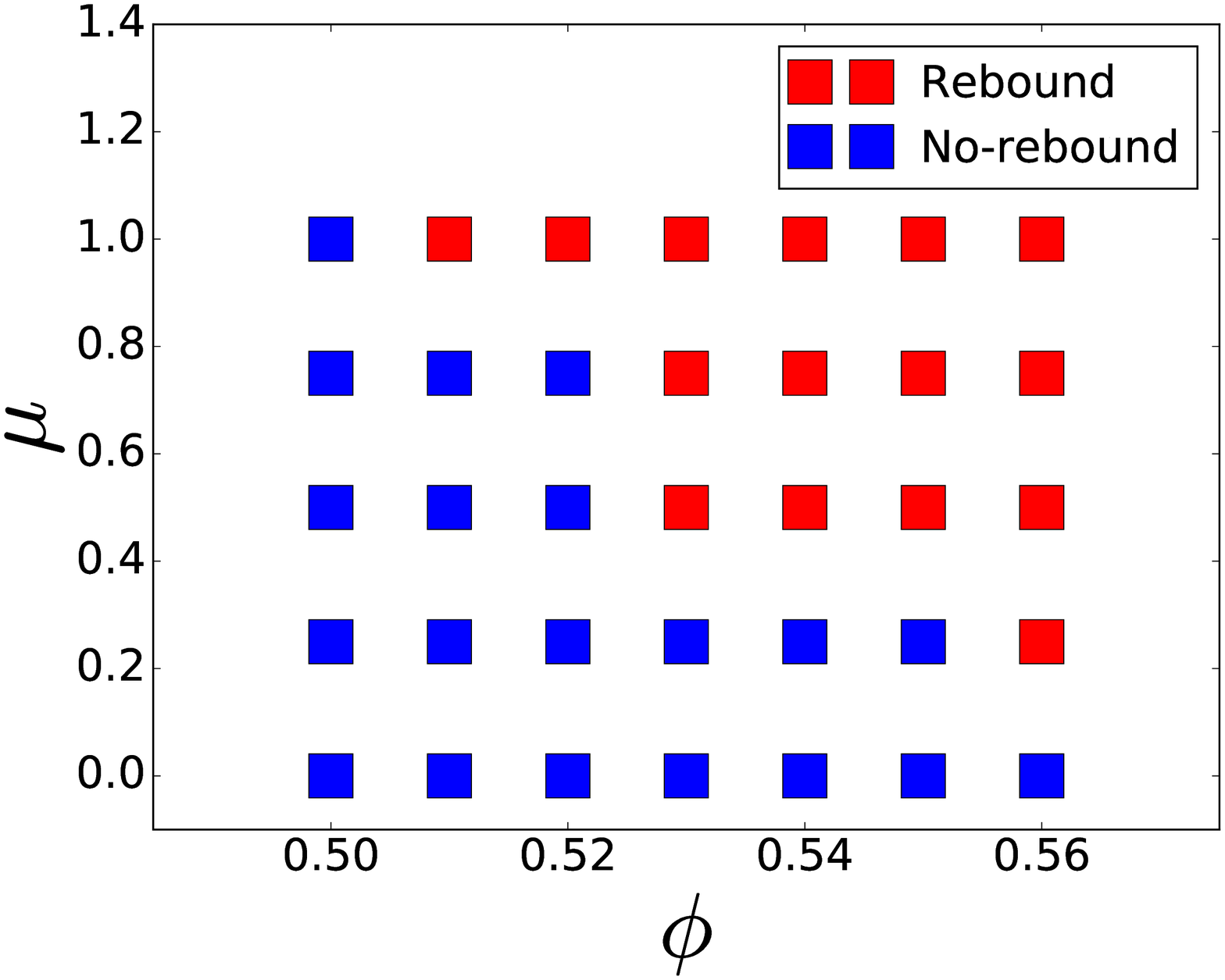}
	 	  \label{fig:6a}
	 }
 	\subfloat[]{
 	 \includegraphics[width=0.3\linewidth]{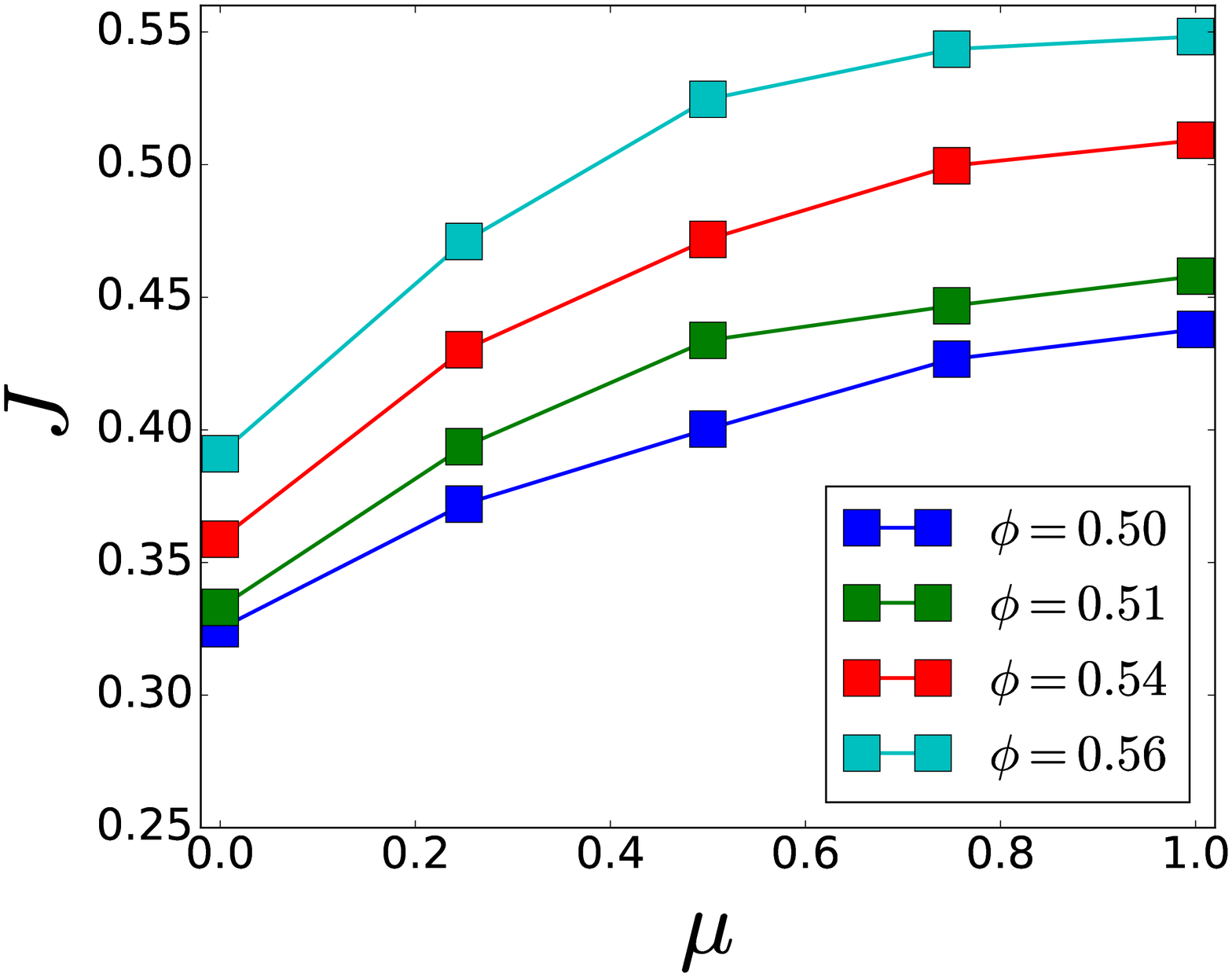}
 	 	  \label{fig:6b}
 	 }
 	\subfloat[]{
 	 \includegraphics[width=0.3\linewidth]{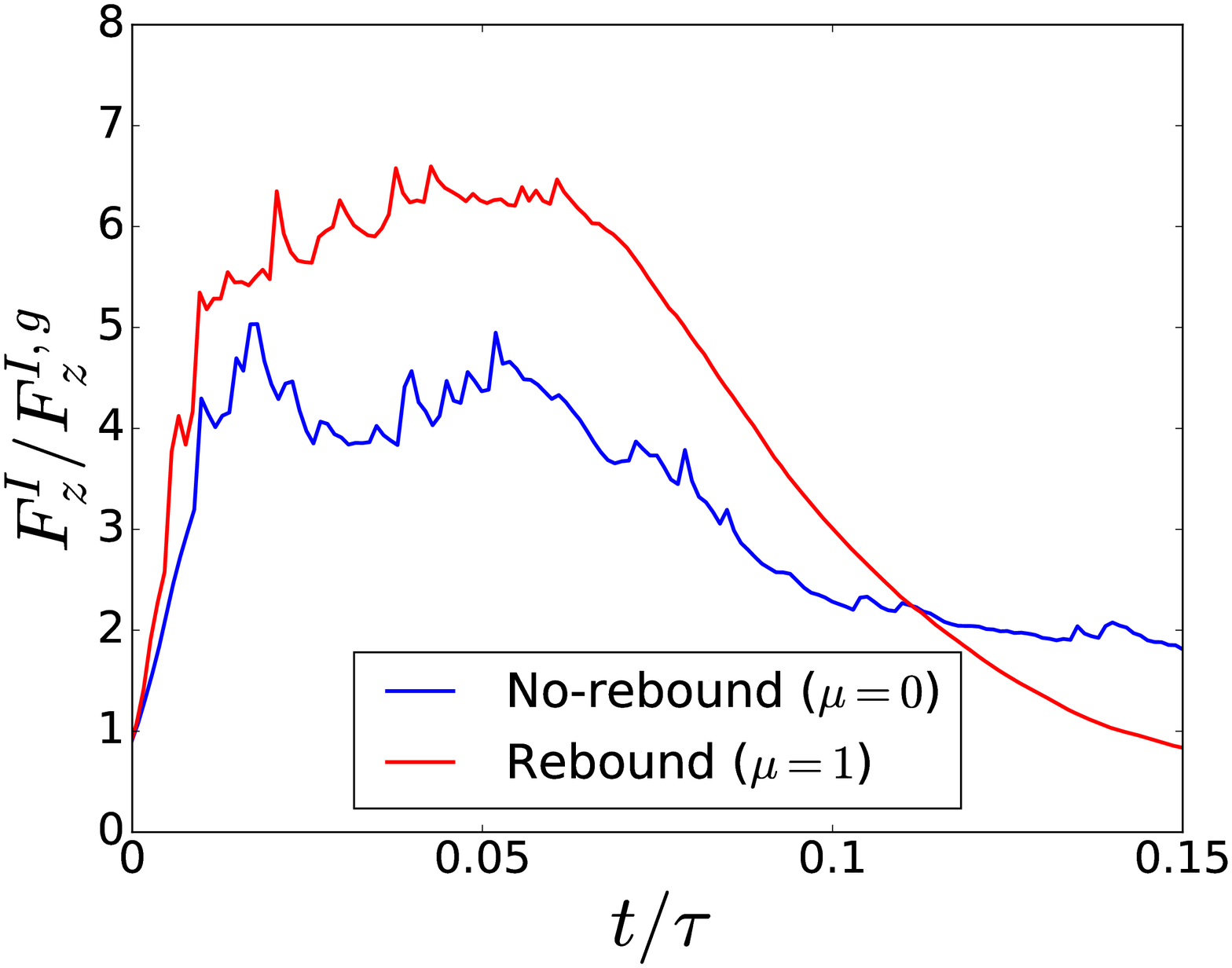}
 	 	  \label{fig:6c}
 	 }
	 \caption{(a) Phase diagram showing whether the impactor rebounds before sinking on a plane of volume fraction $\phi$ and friction coefficient $\mu$. (b) Impulses on the impactor $J$ as functions of the friction coefficient $\mu$ for various $\phi$. (c) Plots of the forces exerted on the impactors for $\phi=0.54$, $u^I_z (t)/u^{*}= 4.26$ for $\mu=0$ (no-rebound) and $\mu=1$ (rebound).}
	\end{figure}

To clarify the roles of mutual friction between particles, we plot the bouncing phase diagram on a plane of the friction coefficient $\mu$ and volume fraction $\phi$ in Fig. \ref{fig:6a}.
Due to the limitation of our computational resources, the data for this phase diagram are obtained by simulations of $N=1200$ particles.
One can verify that the impact-induced hardening is enhanced as $\mu$ increases.
This $\mu$-dependence is analogous to that for DST in dense suspensions under steady shear \cite{otsuki2011,seto2013,mari2014,thomas2018,pradipto2020} and for impact in dry granular materials \cite{kondic2012}. 
To quantify the tendency, we plot the impulse $J$ defined by $J = \int_{t=0}^{t=0.1} F^{I}_{z}(t)dt$ in Fig. \ref{fig:6b}.
One can confirm that the impulse for the frictional case is monotonically increases as the friction coefficient $\mu$ increases for all volume fractions $\phi$.
To investigate how frictional interactions between suspended particles affect the dynamics of the impactor, we plot the time evolution of the forces exerted on the impactor for both the frictional and frictionless cases both for $\phi=0.54$ in Fig. \ref{fig:6c}.
Here, one can verify that the force in the frictional case has a sharper peak than that in the frictionless case. 
The behavior is consistent with Fig. \ref{fig:6b} because the force in the frictional case for $t/\tau < 0.1$ is larger than that in the frictionless case.
It is easy to imagine that the frictional force stabilizes contact points and networks which are needed for rebounds. 
Thus, the friction between particles plays important roles for impact-induced hardening processes. 

\subsection{Local quantities and dynamically jammed region}

\begin{figure*}[htbp]
\centering	
\subfloat[]{\label{fig:7a}%
  \includegraphics[width=0.6\textwidth]{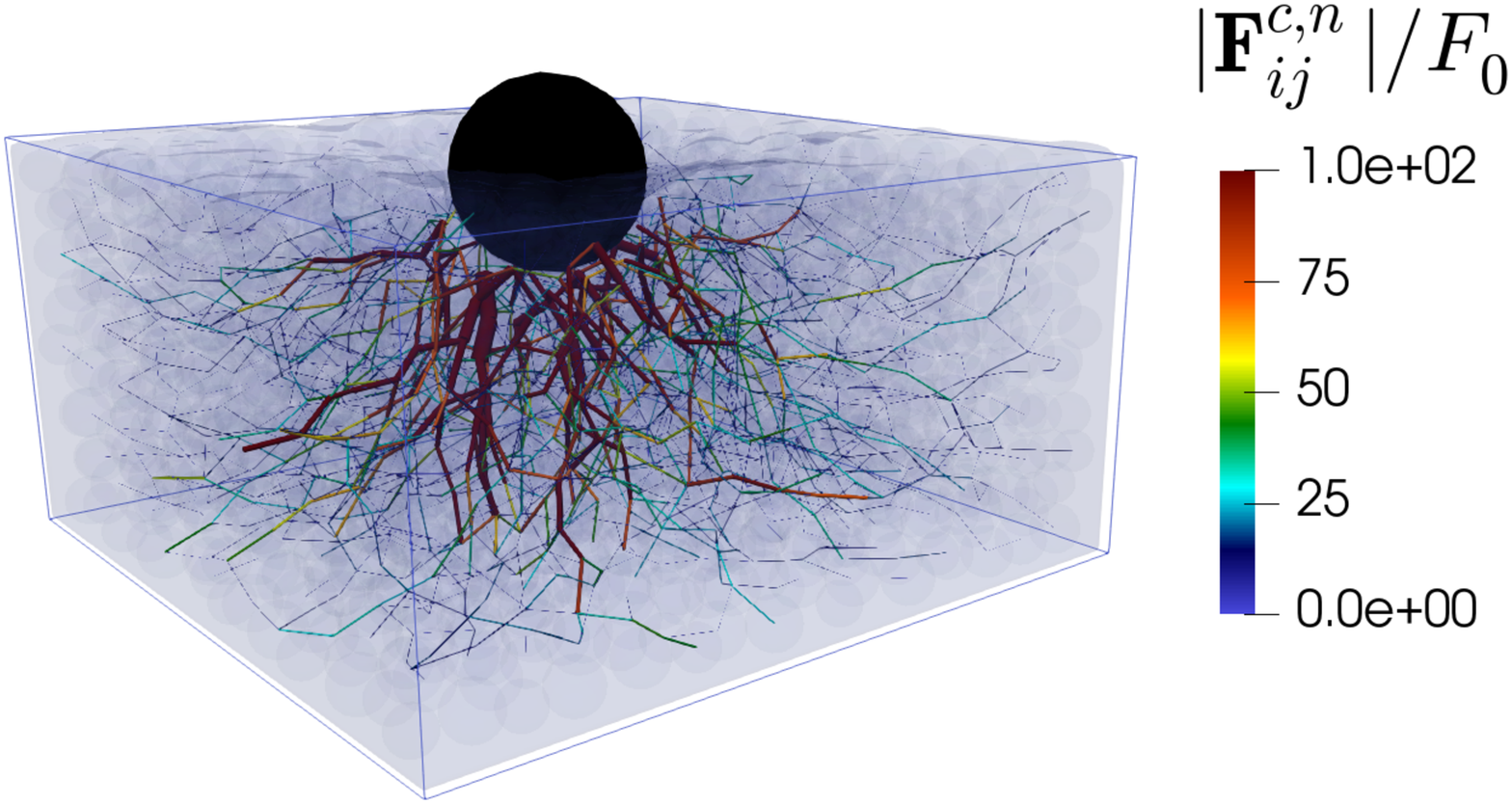}
}

\subfloat[]{\label{fig:7b}%
  \includegraphics[width=0.4\textwidth]{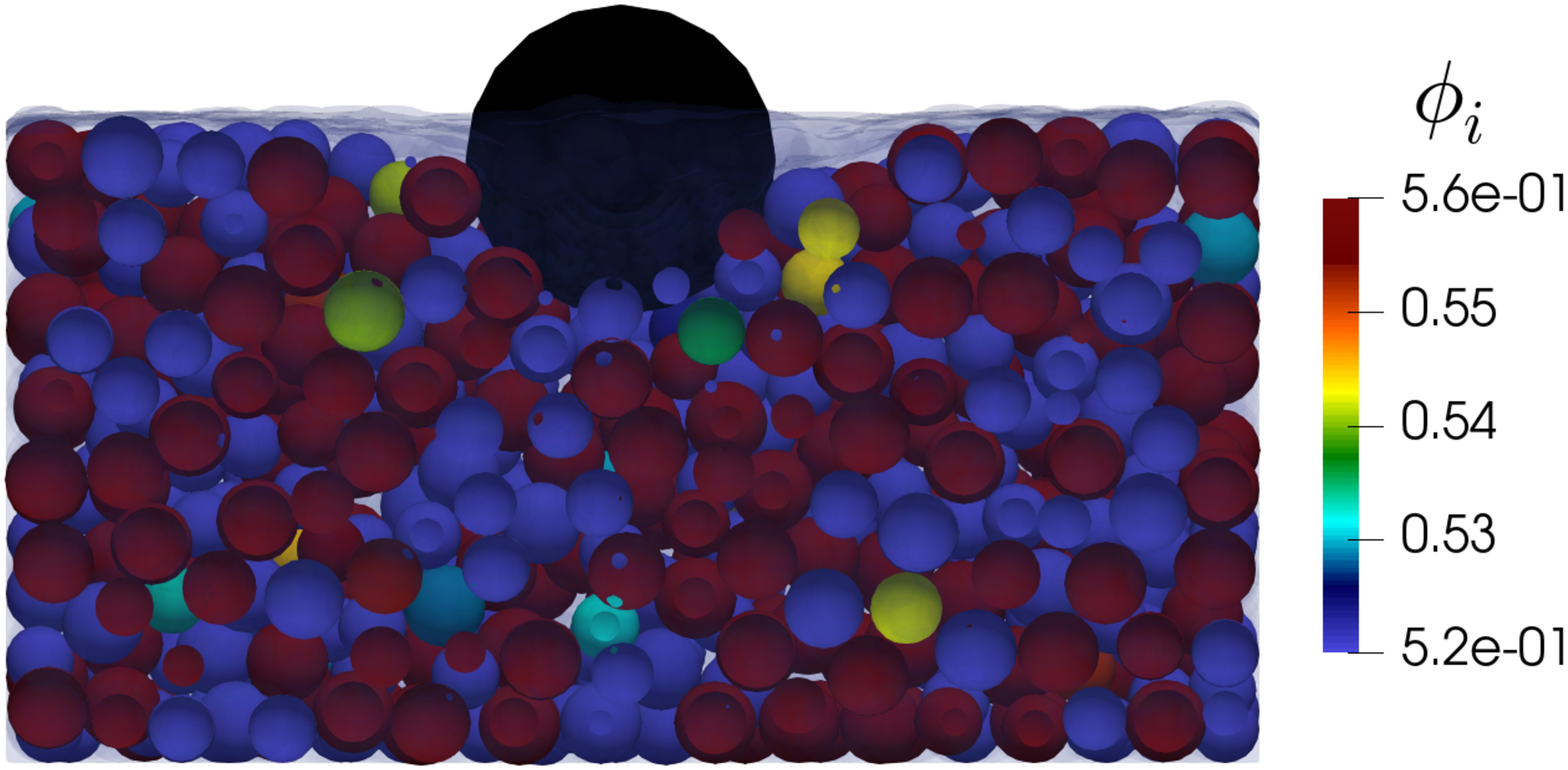}
}
\subfloat[]{\label{fig:7c}%
  \includegraphics[width=0.4\textwidth]{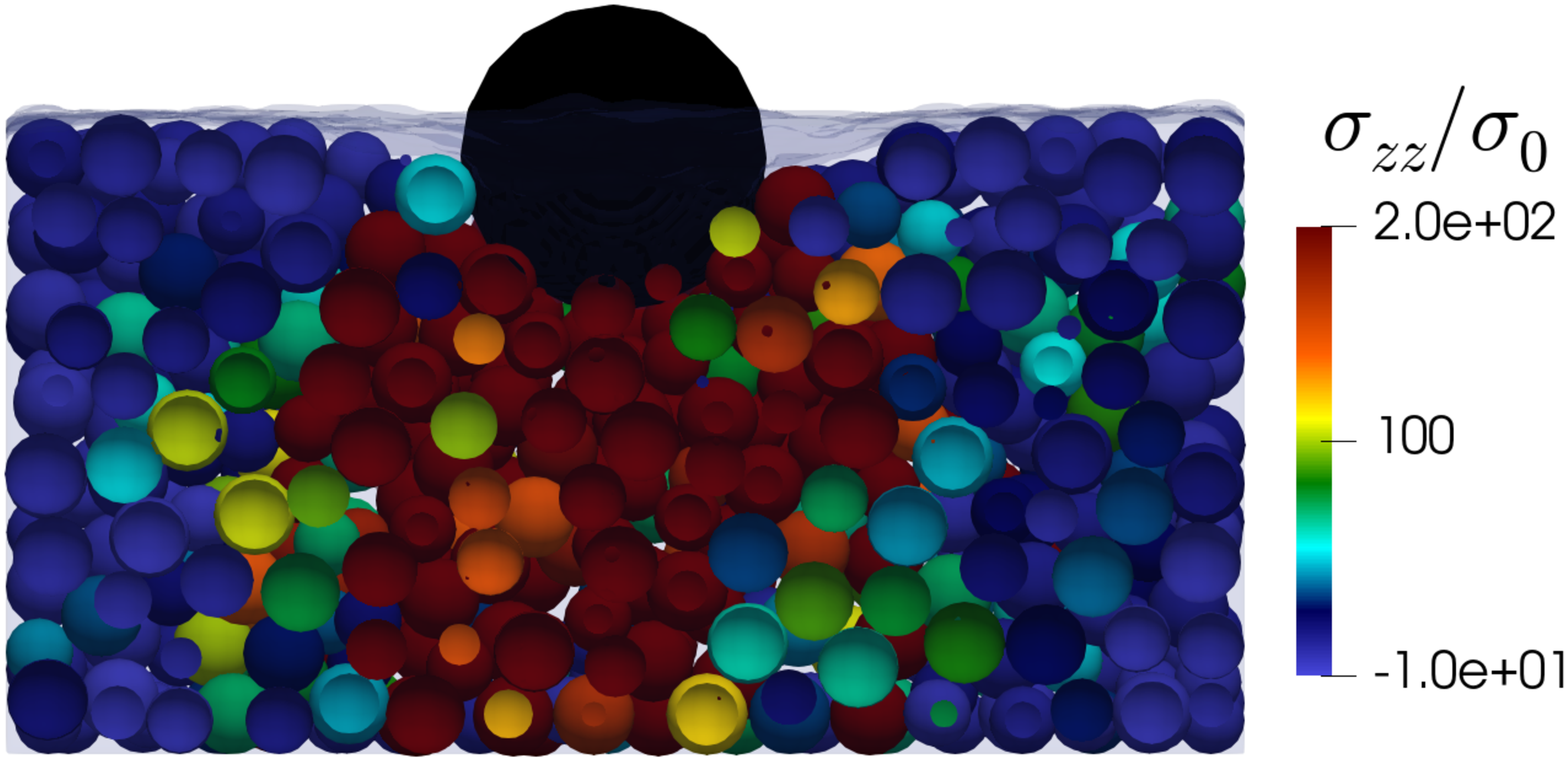}
}

\subfloat[]{\label{fig:7d}%
  \includegraphics[width=0.4\textwidth]{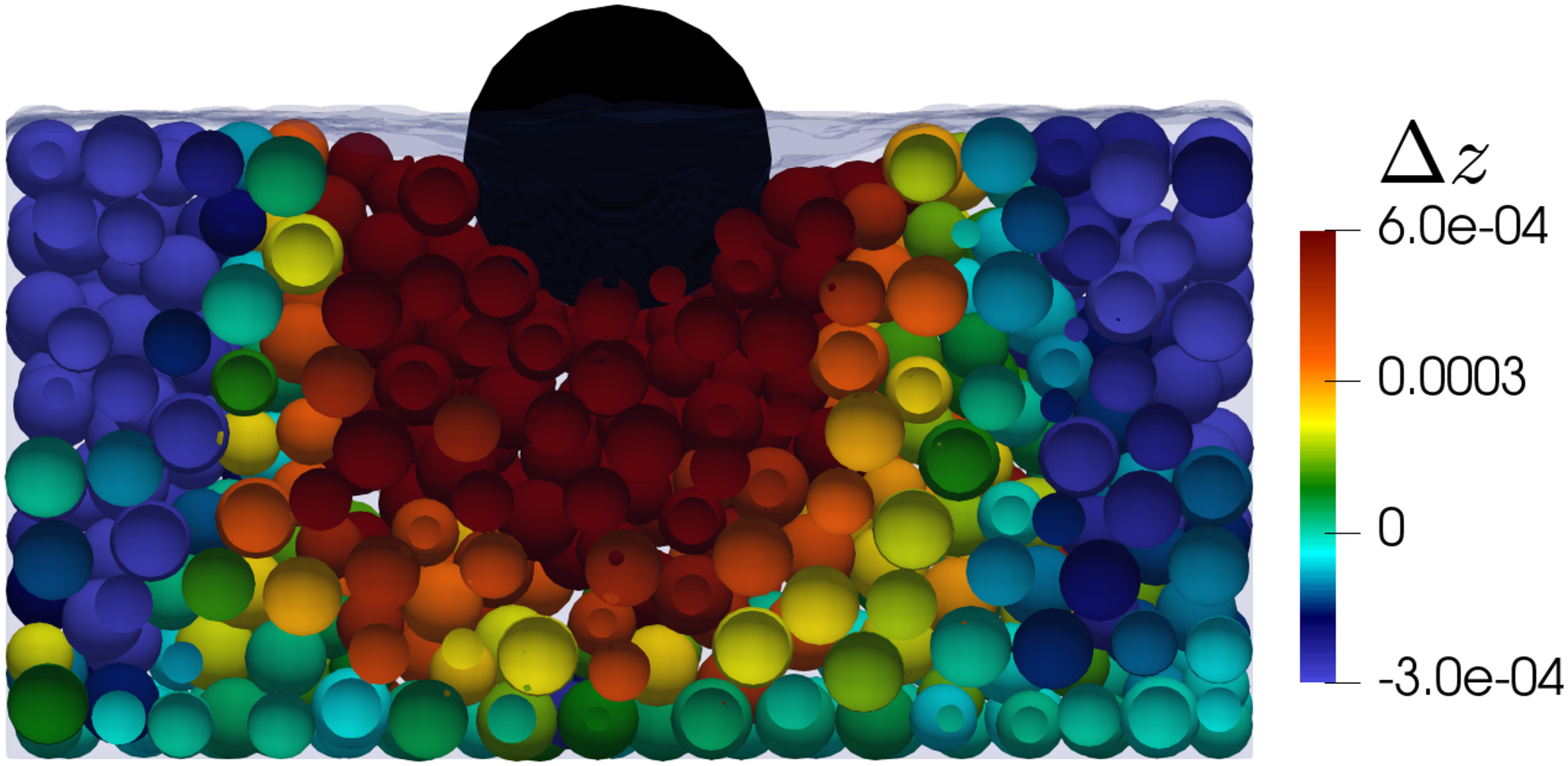}
}
\subfloat[]{\label{fig:7e}%
  \includegraphics[width=0.4\textwidth]{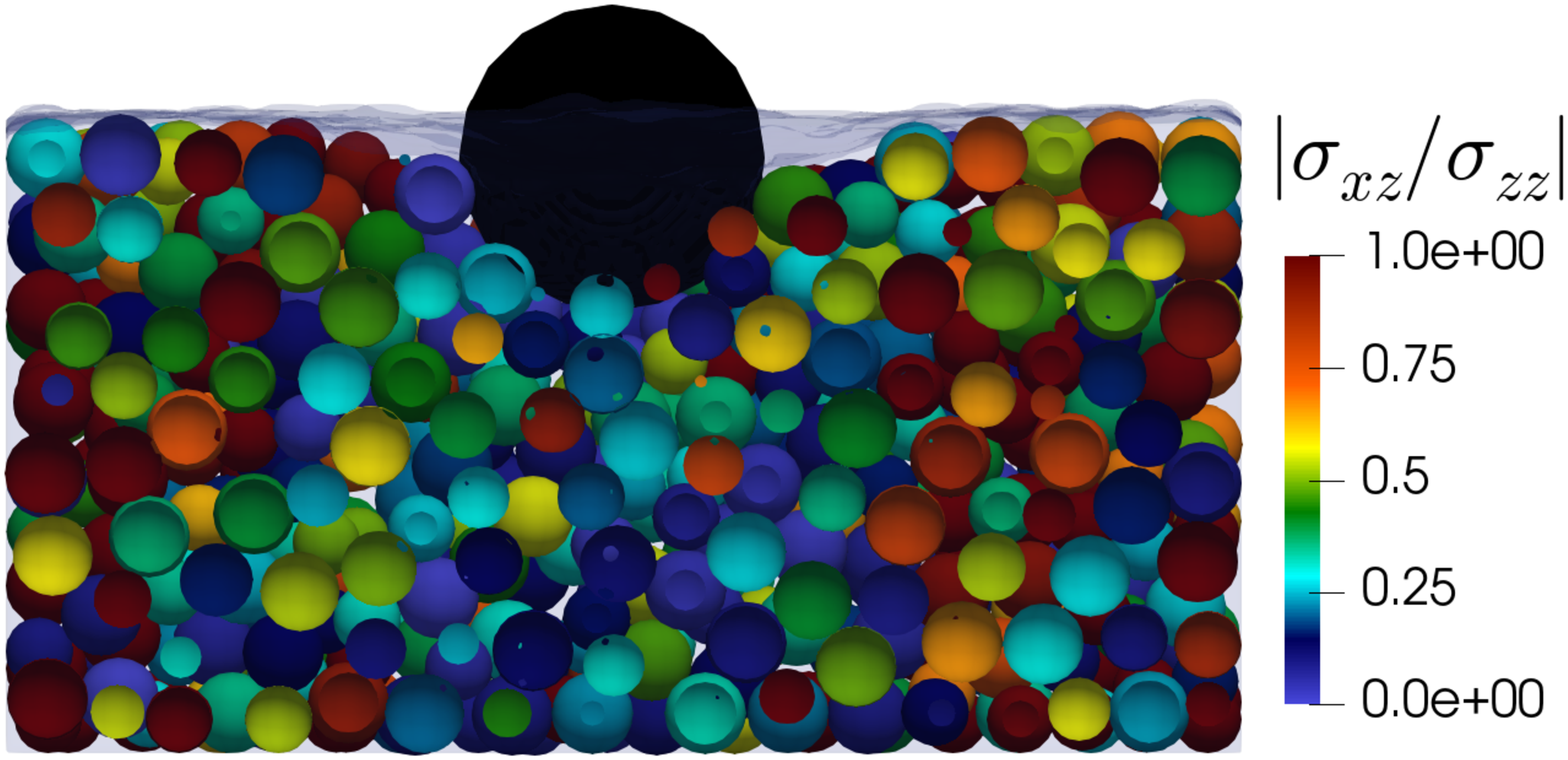}
}

\caption{Visualizations of local quantities for $\phi=0.54$ and $u^{I}_{0,z}/u^{*} = 4.26 $ shortly after the impact ($t/\tau = 0.1$). (a) Force chains of the normal contact forces scaled by the gravitational force $|\bm{F}^{c,n}_{ij}|/F_0$. (b) Local volume fraction $\phi_i$. (c) Magnitude of the dimensionless normal stress $\sigma_{zz}/\sigma_0$. (d) Normal displacement $\Delta z$. (e) Absolute ratio between the shear and normal stress.}
\end{figure*}

To understand the microscopic mechanism behind the impact-induced hardening, we visualize the local responses of the suspension shortly after the impact.
First, we visualize the force chains generated by the impactor by plotting the ratio of the normal contact force to the gravitational force $|\bm{F}^{c,n}_{ij}|/F_0$ in Fig. \ref{fig:7a}.
One can observe the percolating force chains span from the impactor to the boundary without any loops.
The spanned region of force chains from the impactor to the boundaries can be regarded as the dynamically jammed region.
Let us look for quantities to characterize the dynamically jammed region. 
For this purpose, we quantify the local volume fraction $\phi_i$ with the aid of radical Voronoi tesselation \cite{puckett2011} \footnote{We use an open source c++ library Voro++ to construct the radical Voronoi tesselation in our simulation domain \cite{rycroft2009}. Then, the local volume fraction $\phi_i$ is calculated as $\phi_i = v_i / V_i$, where $v_i = 4 \pi a_i^3/3$ is the volume of particle $i$ and $V_i$ is the volume of its corresponding Voronoi cell.}.
In Fig. \ref{fig:7b}, we visualize $\phi_i$ in the sliced region, as shown in Fig. \ref{fig:1c}. 
One can observe that the local volume fraction is almost homogeneous.
We also found that the local volume fraction is not largely affected by the impact \cite{supp_movie2}.
This corresponds to the experimental observation where no detectable increase of packing fraction in the suspension is observed when impact-induced hardening takes place \cite{han2015}.
Thus, the local dense region does not correspond to neither force chains nor the dynamically jammed region.
We visualize the stress $\sigma_{zz}$ on each suspended particle in the sliced region in Fig. \ref{fig:7c}.
Here we observe a localized region with a distinctively high value of $\sigma_{zz}$ corresponding to force chains in Fig. \ref{fig:7a}, which extends from the impactor to the boundary.
In Fig. \ref{fig:7d}, we visualize the particle displacement in normal ($z$-) direction $\Delta z$, also sliced in the middle of the simulation box.
One can observe the existence of a localized region of high normal displacements, which corresponds to the regions in Figs. \ref{fig:7c} and \ref{fig:7a}.
The visualization of $\Delta z$ within our simulation reminisces the experimentally observed one in Refs. \cite{waitukaitis2012, han2015}.
The regions of large $\sigma_{zz}$ (Fig. \ref{fig:7c}), $\Delta z$ (Fig. \ref{fig:7d}), and the force chains (Fig. \ref{fig:7a}) correspond to the dynamically jammed region in Refs. \cite{waitukaitis2012,han2015}.
As indicated in Refs. \cite{waitukaitis2012,han2015,maharjan2018,baumgarten2019}, the propagation speed of the jamming front depends on the impact speed.
After the impactor stops, one can imagine that the vanishing of the stress exerted on the suspension by the impactor allows the suspension to relax and to become soft, which in turn the impactor subsequently sinks after the impact.  
On the other hand, we observe a uniformly weaker magnitude of the shear stress $\sigma_{xz}$ compared to the normal stress $\sigma_{zz}$ as we plot the ratio $\sigma_{xz}/ \sigma_{zz}$ in the sliced region in Fig. \ref{fig:7e}.
Then, the local shear stress is not connected to the dynamically jammed region.
Therefore, this eliminates the conjecture where dynamically jammed region under impact corresponds to the shear jamming and DST.

\subsection{Persistent homology}

To elucidate the role of the force chains in impact-induced hardening, we analyze the topological structure of force chains by using persistent homology analysis \cite{carlsson2009}.
In addition to successfully distinguishing the liquid, amorphous, and crystalline states of, e.g. silicon dioxide \cite{hiraoka2016}, persistent homology allows us to quantify the structure of the force chains in granular materials \cite{kramar2014,takahashi2018} and in dense suspensions under simple shear \cite{gameiro2020}.
Since no persistent loops or higher dimensional structures are observed in the force network in Fig. \ref{fig:7a}, the relevant topological structure is only the connected component represented by the zeroth Betti number $\beta_0$.
On the other hand, the first Betti number $\beta_1$ is important in DST due to the existence of persistent loops in the sheared suspensions \cite{gameiro2020}.

\begin{figure*}[htbp]
\subfloat[]{\label{fig:8a}%
  \includegraphics[width=0.36\textwidth]{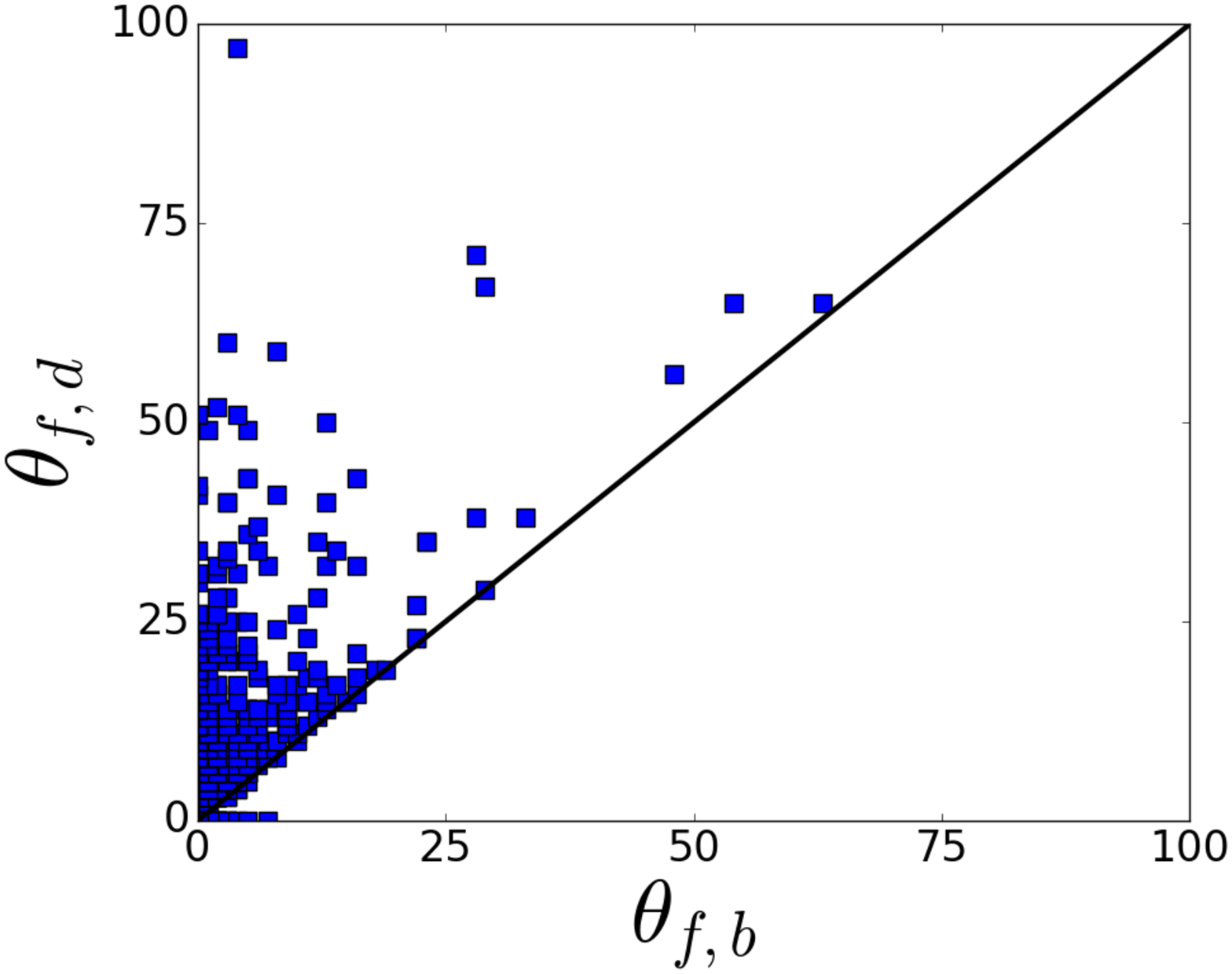}
}
\subfloat[]{\label{fig:8b}%
  \includegraphics[width=0.36\textwidth]{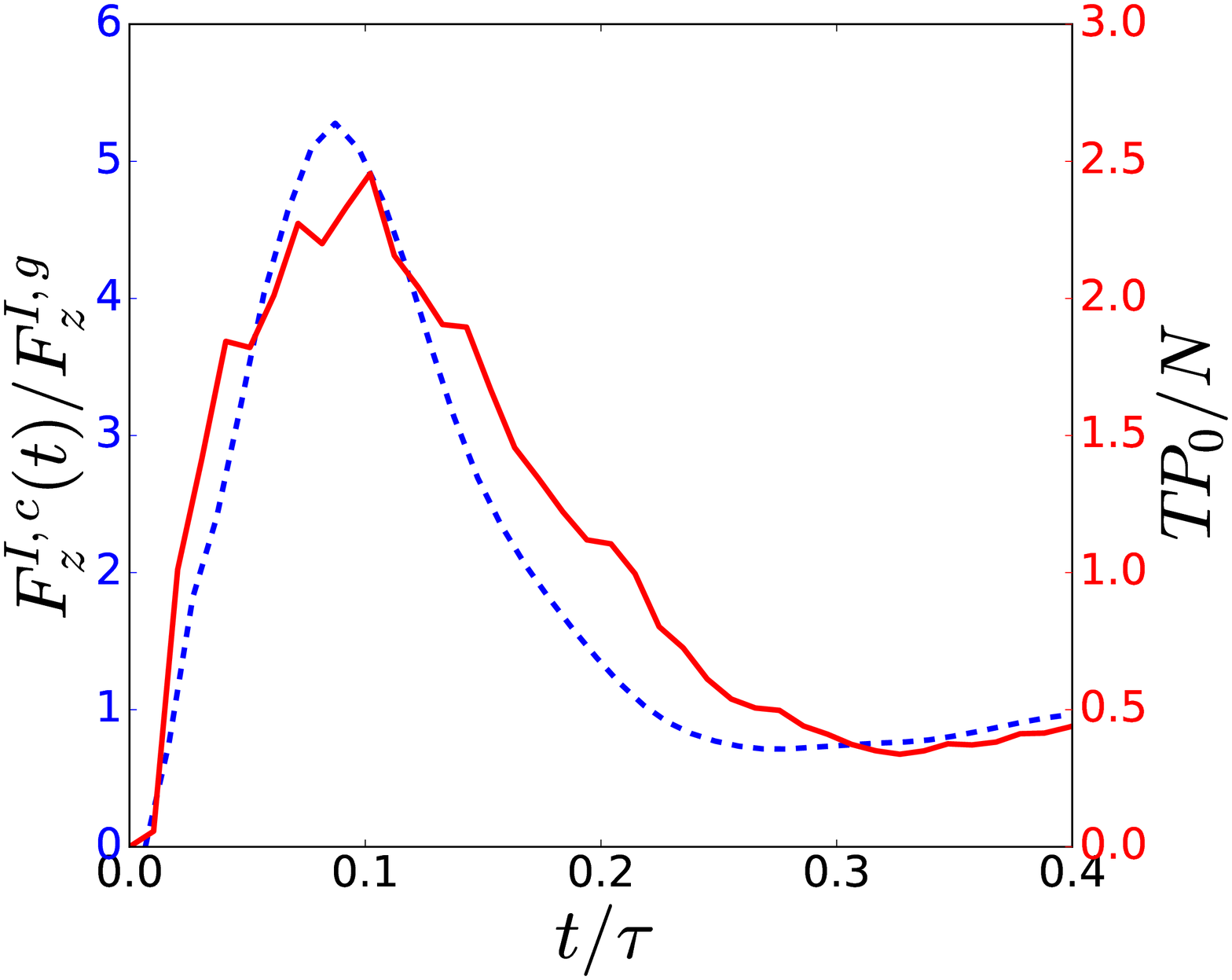}
}
\caption{(a) Persistence diagram of the connected components of force network for $\phi=0.54$, $u^{I}_{0,z}/u^{*} = 4.26 $, and $\mu=1$ shortly after the impact ($t/\tau = 0.1$). (b) Plots of the total persistence of the connected commponents $TP_0$, scaled by the number of suspended particles $N$, against time for $\phi=0.54$ and $u^{I}_{0,z}/u^{*} = 4.26$ (red lines), and the corresponding contact force on the impactor in the $z-$direction $F^{I,c}_{z}$ (dashed blue lines).}
\end{figure*}

The idea of persistent homology is to filter the force chains by increasing threshold $\theta_f$, where a link in a force chain appears when $|\bm{F}^{c,n}_{ij}| / F_0 \leq \theta_f$.
We regard this as the birth of a connected component.
As the threshold further increases, the structure grows in size as additional contacts are added.
When connected components merge, the structure that is born later in the filtration (which has higher birth $\theta_f$) dies.
We record the birth $\theta_f$ as $\theta_{f,b}$ and the death $\theta_f$ as $\theta_{f,d}$.
This rule ensures that $\theta_{f,d} \geq \theta_{f,b}$.
Then, we plot these quantities in the persistence diagram.
In Appendix C, we illustrate the process to translate a force network into a persistence diagram.
The algorithm for filtering chains is available in public domains \cite{mischaikow2013,perseus} \footnote{Note that in Refs. \cite{kramar2014,takahashi2018,gameiro2020}, $\theta_{f,b}$ is always not smaller than $\theta_{f,d}$, since they adopt filtration by reducing the threshold.}.
We plot $\theta_{f,d}$ against $\theta_{f,b}$ for all connected components appearing in Fig. \ref{fig:7a} in the persistence diagram (Fig. \ref{fig:8a}).
The time evolution of the force chains and the persistence diagram can also be seen in Ref. \cite{supp_movie}.
Shortly after the impact, we observe more points far from the diagonal, representing the connected components which persist through the increments of the force threshold with the life span $(\theta_{f,d}-\theta_{f,b})$.
Intuitively, the contact force between particles cannot change abruptly.
Therefore the only possible mechanism for the occurrence of a long lifespan for some connected components is by forming a long chain.
Thus, persistent homology emphasizes the length of the chain instead of its magnitude.
This argument shows that percolated force chains exist.
One point to note is that a component with $\theta_{f,d} = -1$ has infinite persistence, i.e. it does not die until the filtration ends.
The components with infinite persistence represent the contact forces links that do not form any connection with other links.
Persistent homology ignores the effect of such contact forces since we are only interested in extracting the structural information.

The total persistence of the connected components $TP_0$ is the sum of all life spans in the persistence diagram
\begin{equation}
	TP_0 = \sum_{(\theta_{f,d},\theta_{f,b})} (\theta_{f,d}-\theta_{f,b}). \label{eq:tp}
	\end{equation}
This allows us to describe the persistence diagram by a single number.
Higher $TP_0$ means more merging of force chains take place, while $TP_0 = 0$ means that no the connected components are merged.
We plot $TP_0$ scaled by the number of suspended particles $N$ against time in Fig. \ref{fig:8b}.
It is remarkable that $TP_0$ reaches its peak at the same time as the corresponding contact force and that the shape of $TP_0$ is similar to that of the contact force.
Thus, the peak of the contact force inducing the hardening of the suspension originates from the existence of long and sustained force chains.
This can only take place when the force chains are percolated to the boundaries.
Our results provide quantitative proof for the argument in Refs. \cite{waitukaitis2012,maharjan2018,egawa2019} in which the impact-induced hardening takes place when the dynamically jammed region spans from the impactor to the boundary.

To conclude this section, let us re-state the implication from our persistent homology analysis.
First, the magnitude of the force chains is not as important as its topological structure.
Second, persistent homology provides the quantitative proof that the dynamically jammed region which spans from the impactor to the boundary exists.
Third, there are no persistent loops of force chains in our simulations. 
Meanwhile, loops are more significant for sheared suspensions where DST is observed since the total persistence of $\beta_1$ can capture the behavior of the viscosity \cite{gameiro2020}.
This distinction exists because the force chains in sheared suspensions are more structured and uniformly distributed than that in suspensions undergoing impact.
Thus, this gives us another distinction between the impact-induced hardening and the DST or shear jamming.

\section{Conclusions and outlook}

We simulated the impact-induced hardening of suspensions by the LBM simulation with free surface, where the free-falling impactor rebounds for high impact speed with the suspension of high volume fraction involving frictional particles.  
By visualizing each suspended particle, we observed the emergence of the dynamically jammed region with distinctively huge value of normal stress $\sigma_{zz}$, formed by force chains of contacting particles.
Meanwhile, the shear stress $\sigma_{xz}$ of the suspension is not significantly affected by the impact.
We also found that frictional interaction between suspended particles is necessary for the impact-induced hardening to maintain the dynamically jammed region.
The fact that the jammed region is characterized by the normal stress instead of shear stress is important since it distinguishes the impact-induced hardening from shear-induced phenomena such as DST and shear jamming.
Finally, with the aid of persistent homology, (i) we provided the quantitative proof of the existence of a system-spanning dynamically jammed region, (ii) we found that only the topological structure of the force chains is important for the contact force acting on the impactor, and (iii) we did not observe any persistent loops formed by the force networks in contrast to the topological structure for DST. 

Our finding that the response to the impact is affected by the friction coefficient between suspended particles is of interest for future experiments, since one can vary the shape and the asperity of the suspended particles \cite{hsiao2017}.
Some previous papers indicated that the depth of container plays important roles in the impact-induced hardening \cite{roche2013,waitukaitis2012,maharjan2018,mukhopadhyay2018,egawa2019}.
Unfortunately, we presented the results in a container with a fixed depth because of the limitation of our computational resources.
The study on the depth dependence will be reported elsewhere.
In this paper, we only focused on short time behavior after the impact, while a sinking impactor in dense suspensions shows a distinct behavior, as it oscillates and has a stop-go cycle near the bottom of the container \cite{vonkann2011}.
Our simulation will be able to be used to reproduce those results. 
Then, observation of a universal scaling law for impacts on dry granular media was recently reported \cite{krizou2020}.
It would be interesting to investigate whether such a scaling law also exists in the case of impact on dense suspensions.
Finally, a study on the spatial correlation of the force chains has explained the origin of a coherent network that arise from its topological constraints \cite{khrisnaraj2020}.
Such complementary approach will be important to understand the role of the force chains in the impact problem.
These are targets of our next research.

\begin{acknowledgments}
One of the authors (HH) thanks H. Katsuragi for his insights in the early stages of this research project.
One of the authors (P) expresses his gratitude to A. Leonardi for sharing his lattice Boltzmann code.
The authors thank M. Otsuki, R. Seto, T. Nakamura, and T. Yamaguchi for their useful comments and fruitful discussions.
The authors also thank V. M. M. Paasonen for his critical reading of this manuscript.
All numerical calculations were carried out at the Yukawa Institute for Theoretical Physics (YITP) Computer Facilities, Kyoto University, Japan.
This work was partially supported by ISHIZUE 2020 of Kyoto University Research Development Program.
\end{acknowledgments}

\appendix

\section{Lattice Boltzmann method for suspensions with free surface}

\subsection{Review on the lattice Boltzmann method}

In this section, we review the lattice Boltzmann method (LBM) based on Refs. \cite{pradipto2019thesis,pradipto2020,ladd1994a,ladd1994b,leonardi2015thesis,succi2001}.
Due to the discrete nature of the LBM, one needs to discretize the unit of length into the lattice unit $\Delta x$. 
We take the lattice unit $\Delta x$ as $\Delta x = 0.2 a_{\text{min}}$ ($a_{\text{min}}$ is the radius of the smallest particle).
In LBM, the hydrodynamic fields (density $\rho_f$ and velocity $\bm{u}_f$) are calculated on nodes $\bm{r}$ inside cells of a fixed Cartesian grid as
\begin{align}
	\rho_f (\bm{r})  = \sum_{\bm{q}} f_{\bm{q}} (\bm{r}) \Delta c^3, \qquad  \rho_f \bm{u}_f (\bm{r}) = \sum_{\bm{q}} f_{\bm{q}} \bm{c}_{\bm{q}} (\bm{r}) \Delta c^3, \label{eq:1}
	\end{align}
where $\bm{c}_{\bm{q}}$ is the lattice velocity of the direction $\bm{q}$, and $\Delta c^3$ is the volume element in the velocity space with $\Delta c = \Delta x / \Delta t$.
$f_{\bm{q}} (\bm{r})$ is the abbreviation of $f_{\bm{q}} (\bm{r},t)$ which is the discrete distribution function and has the dimension of mass density. 
The evolution equation for $f_{\bm{q}} (\bm{r}, t)$ is 
\begin{align}
	f_{\bm{q}} (\bm{r} + \bm{c}_{\bm{q}} \Delta t, t + \Delta t) = f_{\bm{q}} (\bm{r},t) + \Delta t ( \Omega_{\bm{q},c}+ \Omega_{\bm{q},f}), \label{eq:2}
	\end{align}
where $\Omega_{\bm{q},c}$ is the collision operator and $\Omega_{\bm{q},f}$ is an additional operator if a volumetric force density $\tilde{\bm{f}}$ acts on the system.
We use the Bhatnagar-Gross-Krook approximation for the collision operator \cite{bhatnagar1954}, which relaxes the system to the equilibrium state $f_{\bm{q}}^{\text{eq}}$ as
\begin{equation}
	\Omega_{\bm{q},c} = \frac{f_{\bm{q}}^{\text{eq}} - f_{\bm{q}}}{\tau_{r}}, \label{eq:3}
	\end{equation}	
where $\tau_r$ is the relaxation time relating to the kinematic viscosity $\nu$ as $\tau_r =  \Delta t/2 + \nu / c_s^2$, with the lattice sound speed $c_s = \sqrt{\Delta c^2/3}$.	
The equilibrium distribution function $f_{\bm{q}}^{\text{eq}}$ is calculated as
\begin{equation}
	 f_{\bm{q}}^{\text{eq}} \Delta c^3= w_{\bm{q}} \rho_f \bigg[1 + \frac{\bm{c}_{\bm{q}} \cdot \bm{u}_f}{c_s^2} + \frac{(\bm{u}_f \bm{u}_f :( \bm{c}_{\bm{q}} \bm{c}_{\bm{q}} - c_s^2 \bm{I} ))}{2 c_s^4} \bigg],
	\end{equation}
where $w_{\bm{q}}$ is the lattice weight that depends on the configurations.	
For $\Omega_{\bm{q},f}$, we employ \cite{guo2002}
\begin{equation}
	\Omega_{\bm{q},f}  \Delta c^3= w_{\bm{q}} \bigg(1 - \frac{\Delta t}{2\tau_r} \bigg) \bigg[ \frac{(\bm{c}_{\bm{q}} - \bm{u}_f)}{c_s^2} +  \frac{( \bm{c}_{\bm{q}} \cdot \bm{u}_f)}{c_s^4} \bm{c}_{\bm{q}} \bigg] \cdot \tilde{\bm{f}}. \label{eq:guo}
	\end{equation}
As a result, the macroscopic velocity is changed so the second term in Eq. \eqref{eq:1} becomes
	\begin{align}
 \rho_f  \bm{u}_f(\bm{r}) = \sum_{\bm{q}}  \left\{ f_{\bm{q}}  \bm{c}_{\bm{q}} (\bm{r}) \Delta c^3 + \frac{\Delta t \tilde{\bm{f}} (\bm{r}) }{2}\right\}. \label{eq:1b}
		\end{align}

\subsection{Handling the free surface of the fluid}
\begin{figure}[htbp]
 	\includegraphics[width=0.3\linewidth]{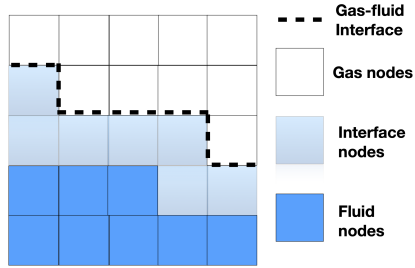}
 	\caption{Illustration of the the division of the lattice nodes into fluid, interface and gas nodes.}
	\label{fig:9}
 	\end{figure}

To simulate the free surface, we need to implement the mass tracking algorithm \cite{korner2005,svec2012,leonardi2015}.
First, we assign a type of nodes such as the fluid, interface, or gas node for each node, where the interface node exists between the fluid and gas nodes as in Fig. \ref{fig:9}.
Note that Eqs. \eqref{eq:1} and \eqref{eq:2} are only used in the fluid and interface nodes.

A gas node represents the cell which is not occupied by the fluid, hence $f_{\bm{q}} = 0$.
An interface node expresses the interface between the fluid and gas, where the streaming and collision of $f_{\bm{q}}$ exists as in fluid nodes.
Here, we introduce a variable $m_f$, which represents the density of the fluid in a single cell, to track the evolution of the surface.
The interface node turns into a fluid node if $m_f \geq \rho^{*}_f$ or into a gas node if $m_f \leq 0$, where $\rho^{*}_f$ is the unit density of the fluid.
Therefore, the state of each node is characterized by the liquid fraction $\lambda$:
\begin{equation}   
     \begin{cases}
        \lambda = 1&\quad \text{if the node is liquid}\\
		0 < \lambda < 1 &\quad \text{if the node is interface,}\\
		\lambda = 0 &\quad \text{if the node is gas,}\\
     \end{cases}
	 \end{equation}
where $m_f = \lambda \rho_f$.
The evolution of the $m_f$ is determined by the balance between the populations streaming into the node $f_{\bm{q}'}(\bm{r} + \bm{c}_{\bm{q}'}\Delta t,t)$ ($\bm{q}' = -\bm{q}$) and out of the node $f_{\bm{q}} (\bm{r},t)$
\begin{align}
	m_f (t + \Delta t) =\Delta t \sum_{\bm{q}} \alpha_{\bm{q}} (f_{\bm{q}'}(\bm{r} + \bm{c}_{\bm{q}'}\Delta t,t) - f_{\bm{q}} (\bm{r},t)) \Delta c^3  +  m_f(t), \label{eq:massevol}
	\end{align}
where $\alpha_{\bm{q}}$ is a function of $\lambda$ of the neighboring node (located at $\bm{r} + \bm{c}_{\bm{q}'}\Delta t$).

\begin{equation}   
\alpha_{\bm{q}} = 
     \begin{cases}
        \frac{1}{2}[\lambda (\bm{r},t) + \lambda (\bm{r}+\bm{c}_{\bm{q}'}\Delta t,t)] &\quad \text{if } f_{\bm{q}'}(\bm{r} + \bm{c}_{\bm{q}'}\Delta t,t) \text{ streams from an interface node},\\
		1 &\quad \text{if } f_{\bm{q}'}(\bm{r} + \bm{c}_{\bm{q}'}\Delta t,t) \text{ streams from a fluid node},\\
		0 &\quad \text{if } f_{\bm{q}'}(\bm{r} + \bm{c}_{\bm{q}'}\Delta t,t) \text{ streams from a gas node}.\\
     \end{cases}
\end{equation}

When an interface node turns into a fluid node, the neighboring gas nodes turn into interface nodes.
When an interface node turns into a gas node, the neighboring fluid nodes turn into interface nodes.
Although the density in a continuum model must be conserved, the discrete model can contain small loss or gain of $m_f$.
The surplus (or shortfall, including the possibility of negative density) of $m_f$ is then computed at every time step and is corrected to satisfy the conservation among all interface nodes.

\textit{Fixed-pressure boundary condition:}
As stated before, LBM equations are solved only in the liquid and interface nodes.
This creates a problem in the implementation since the population streaming to the interface nodes from gas nodes which is necessary in Eq. \eqref{eq:2} is not well-defined.
Assuming that the gas nodes are always in equlibrium characterized by $f_{\bm{q}}^{\text{eq}}$ and have the same $\bm{u}^{\text{in}}_{f}$ and $c_s$ as the interface nodes, with a constant atmospheric density $\rho_{a} = 0.3 \rho_f$, we can solve Eq. \eqref{eq:2}) \cite{korner2005, leonardi2015thesis}, 
This is analogous to applying a fixed-pressure boundary condition at the interface and local symmetry conditions for the velocity.
The condition $\rho_a > 0$ also gives an effective surface tension to the system \cite{korner2005}.

\subsection{Solid boundaries and the fluid-particle coupling}
We implement two coupling schemes to handle solid boundaries within our simulations.
We use the bounce-back rules for no-slip boundary condition on walls and the surface of the impactor, while we use the direct forcing scheme for suspended particles.
 
\textit{The bounce-back rule} simply states that whenever a population is streaming towards a wall, this population is reflected and bounced back in the opposite direction.
This rule can be expressed as 
\begin{equation}
	f_{\bm{q}'} (\bm{r},t+\Delta t) = f_{\bm{q}} (\bm{r},t),
	\end{equation}
in LBM notation.	
If the wall is moving, the reflection has to take into account the momentum transfer by an addititonal term \cite{ladd1994a,ladd1994b}
\begin{equation}
	\{f_{\bm{q}} (\bm{r},t) - f_{\bm{q}'} (\bm{r},t+\Delta t)\}\Delta c^3 =  \bigg(  \frac{2 w_{\bm{q}} \rho_f \bm{u}_w \cdot \bm{c}_{\bm{q}}}{c_s^2} \bigg), \label{eq:ladd}
		\end{equation}	
where $\bm{u}_w$ is the wall velocity.
Here, $\bm{u}_w$ is calculated as
\begin{equation}
	\bm{u}_w (\bm{r}) = \bm{u}^{I} + (\bm{r} - \bm{R}^I) \times \bm{\omega}^{I},
	\end{equation}
where $\bm{u}^I$ and $\bm{\omega}^I$ are the translational velocity and the angular velocity on the surface of the impactor, respectively, and $\bm{R}^I$ denotes the center of mass of the impactor.
The momentum exchange described in Eq. \eqref{eq:ladd} results in a force on each node on the impactor surface $\bm{\tilde{F}}(\bm{r})$ as 
\begin{equation}
	\bm{\tilde{F}} (\bm{r}) = \frac{\Delta x^3}{\Delta t} \bigg(2 f_{\bm{q}} (\bm{r},t) \Delta c^3 - \frac{2 w_{\bm{q}} \rho_f \bm{u}_w \cdot \bm{c}_{\bm{q}}}{c_s^2} \bigg) \bm{c}_{\bm{q}}.
	\end{equation}
The hydrodynamic force on the impactor $\bm{F}^{I,h}$ is the sum of the forces for all nodes in the surface as $\bm{F}^{I,h} = \sum_{\bm{r} \in \text{surface}} \bm{\tilde{F}} (\bm{r})$, while $\bm{T}^{I,h} = \sum_{\bm{r} \in \text{surface}} (\bm{r} - \bm{R}^I) \times \bm{\tilde{F}} (\bm{r})$ is the hydrodynamic torque.
   
\textit{Direct forcing:} By using the immersed boundary method, we calculate the hydrodynamic force through an additional discretization of particles into a set of segments $\bm{r}_{\text{cell}}$.
These particle segments are related to the fluid simulation by an interpolating function \cite{peskin1972}.
We implement the simplified version \cite{leonardi2015,leonardi2015thesis}, where the segments correspond to the lattice nodes of the LBM $\bm{r}_{\text{cell}} = \bm{r}$.
Since the volume of a cubic cell is unity, the hydrodynamic force on each cell $\bm{\tilde{F}}_{\text{cell}} (\bm{r})$ can be computed directly from the velocity differences
\begin{equation}
	\bm{\tilde{F}}_{\text{cell}} (\bm{r}) = \frac{\Delta x^3}{\Delta t}  \rho_f (\bm{r}) [\bm{u}_f (\bm{r}) - \bm{u}_{\text{cell}} (\bm{r})],
	\end{equation}
where $\bm{u}_{\text{cell}}$ is the velocity of the particle cell
	\begin{equation}
		\bm{u}_{\text{cell}} (\bm{r}) = \bm{u} + (\bm{r} - \bm{R}) \times \bm{\omega},
		\end{equation}
where $\bm{u}$, $\bm{R}$, and $\bm{\omega}$ are the translational velocity, center of mass,	and angular velocity of the suspended particles, respectively.	
The resultant hydrodynamic force on each suspended particle $\bm{F}^h$ is the sum of all forces on the cells inside the particle $l$ as $\bm{F}^h = \sum_{\bm{r} \in l} \bm{\tilde{F}}_{\text{cell}} (\bm{r})$.
Similarly, the torque is given by $\bm{T}^{h} = \sum_{\bm{r} \in l} (\bm{r} - \bm{R}) \times \bm{\tilde{F}}_{\text{cell}} (\bm{r})$.
Note that this method requires a contribution to the body force density of the fluid $\tilde{\bm{f}}$.
Therefore, we calculate $\tilde{\bm{f}}$ in  Eqs. \eqref{eq:1b} and \eqref{eq:guo} as
\begin{equation}
\tilde{\bm{f}} (\bm{r}) = - \rho_f g \hat{\bm{z}} - \frac{\bm{\tilde{F}}_{\text{cell}} (\bm{r})}{\Delta x^3}. \label{eq:bodyforce2}
\end{equation}
The first term in Eq. \eqref{eq:bodyforce2} comes from gravity.
Note that this scheme requires all segments of each particle to be inside the solvent fluid.

\section{Finite size effects}

\begin{figure}[htbp]
 	\includegraphics[width=0.36\linewidth]{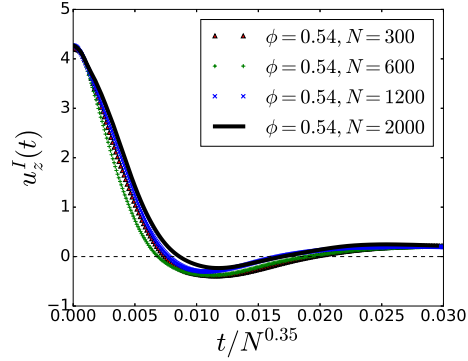}
 	\caption{Plots of impactor speeds in the $z$-direction, $u^I_z (t)/u^{*}$, against time for several numbers of particles $N$ for $\phi=0.54$. The time scale $t$ is scaled by the particle numbers $N^{\alpha}$ with $\alpha=0.35$}
	\label{fig:10}
 	\end{figure}

In this appendix, we examine how the impactor dynamics depends on the number of particles in our system.
We plot the time evolution of the impactor velocity for several numbers of particles $N$ for $\phi = 0.54$ and $u^{I}_{0,z}/u^{*} = 4.26$ in Fig. \ref{fig:10}.
As mentioned in Sec. II, we keep the ratios of the impactor radius to the width and depth of the box as $W/a_I = 8$, $D/a_I = 8$, and $H/a_I = 4$.
Therefore, varying the numbers of particles $N$ also changes the ratio of impactor radius $a_I$ to the smallest suspended particles radius $a_{\rm min}$ as $a_I/a_{\rm min} = 2.25, 3, 3.75,$ and $4.5$ for $N=300,600,1200,$ and $2000$, respectively.

We found that the system size dependence mainly appears as the time scale of the impact processes. 
Although the impact velocity depends a little on the system size (and as a result, the phase diagrams also depend a little on the system size), such system size dependences are not significant.
For instance, if we scale the time by $N^{\alpha}$ with exponent $\alpha=0.35$ for the data of $\phi=0.54$, we can obtain an approximate universal curve of the impact speed.
Thus, one can guess the behavior in the thermodynamic limit from the simulations with small systems. 
Note that the value of $\alpha$ might depend on $\phi$.
A systematic study in finite-size scaling for simulations of dense suspensions under impact will be reported elsewhere.

\section{Brief explanation of persistent homology}
\begin{figure*}[htbp]
\subfloat[]{\label{fig:11a}%
  \includegraphics[width=0.33\textwidth]{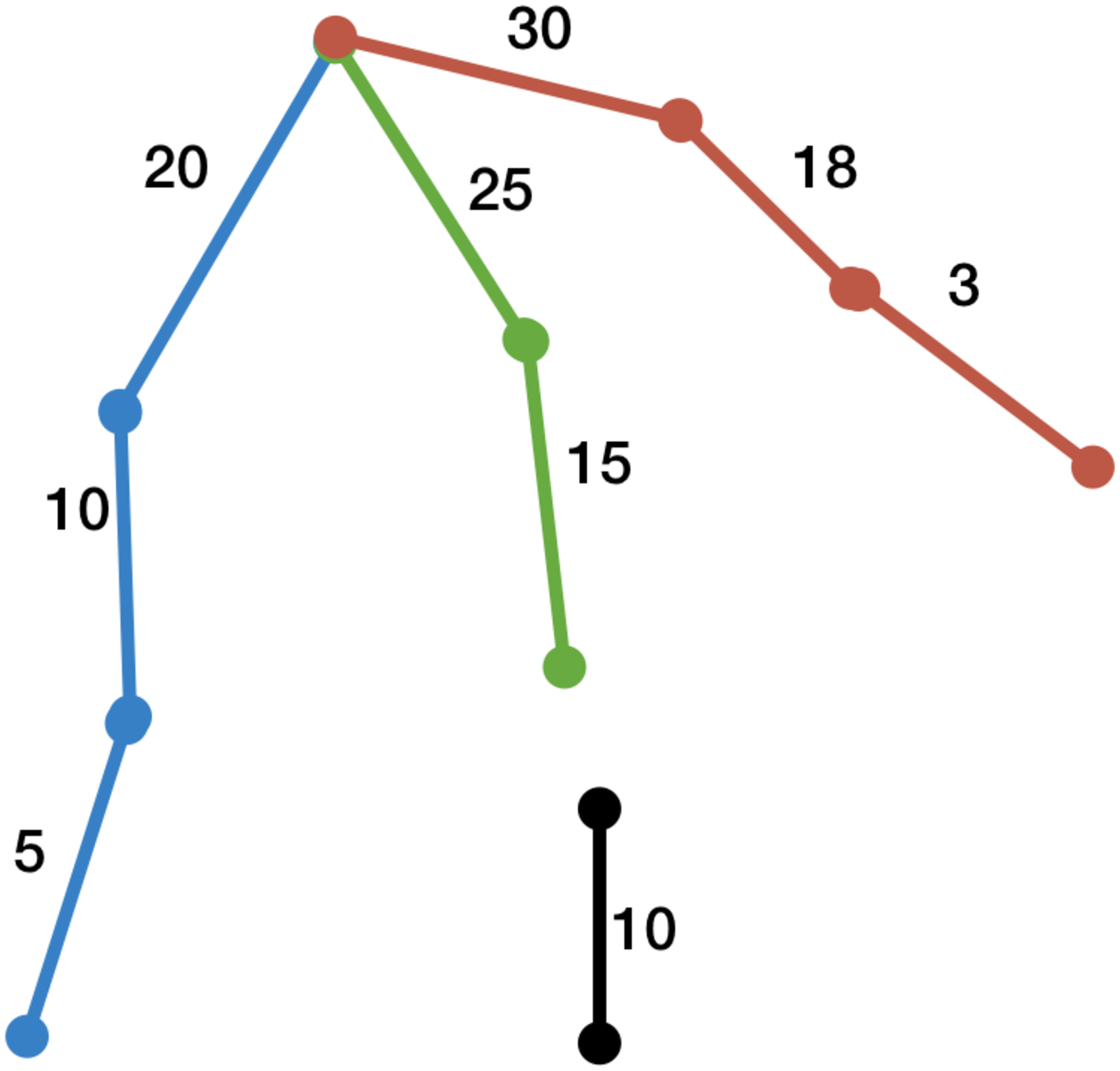}
}
\subfloat[]{\label{fig:11b}%
  \includegraphics[width=0.33\textwidth]{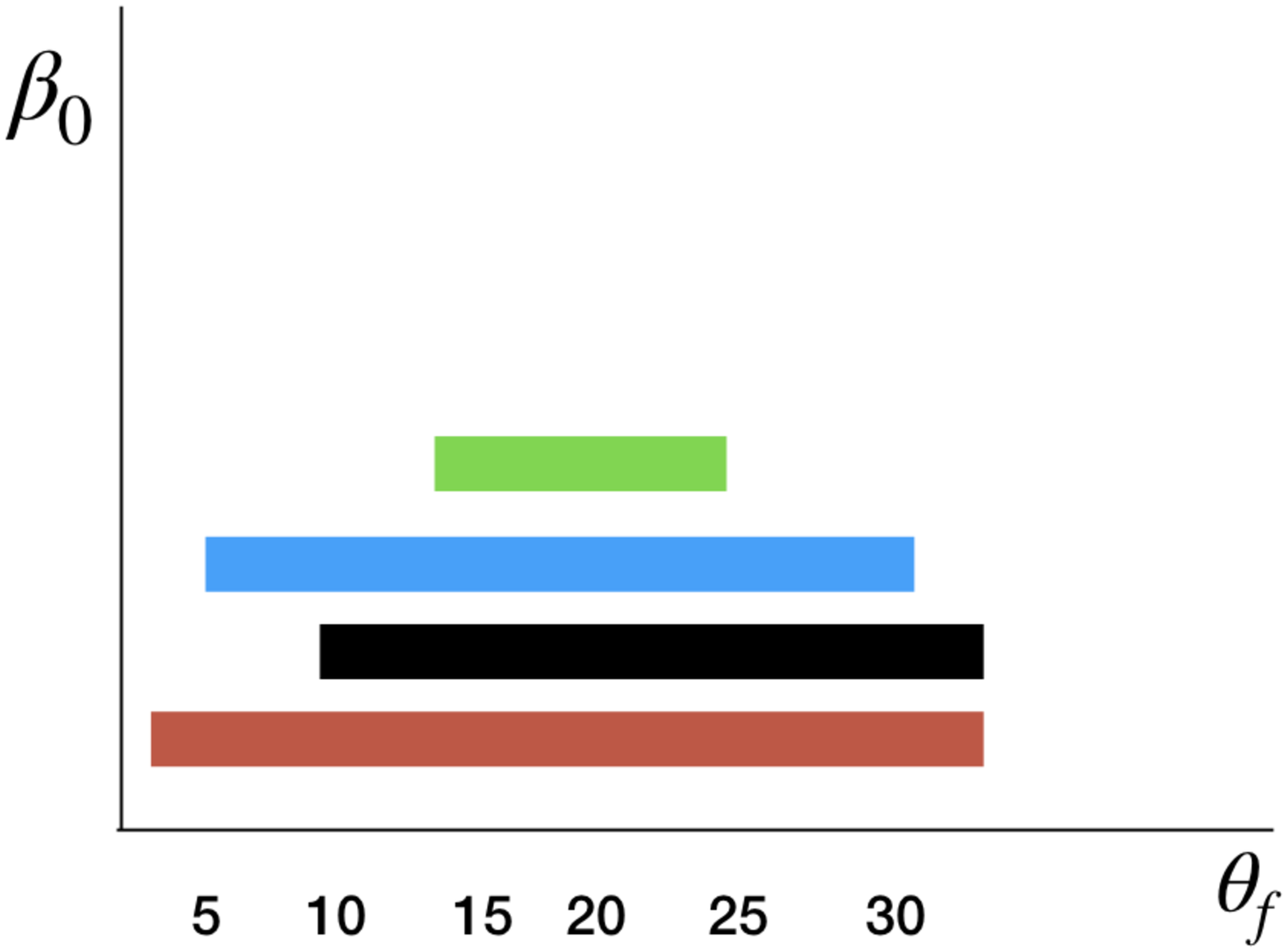}
}
\subfloat[]{\label{fig:11c}%
  \includegraphics[width=0.33\textwidth]{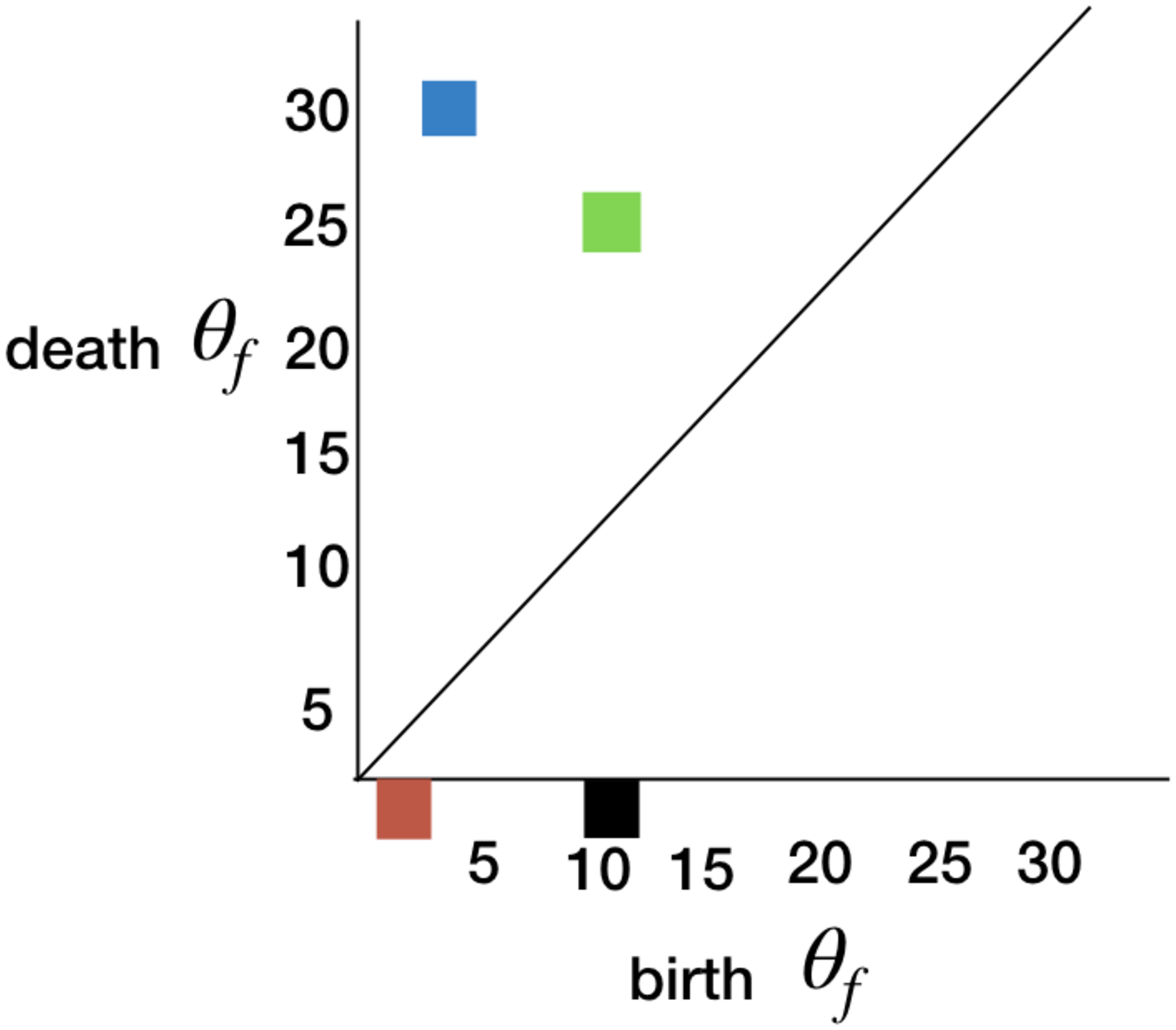}
}
\caption{(a) An illustration of a force network configuration, where the numbers represent the force magnitude and the colors represent each connected component. (b) The corresponding barcode and (c) the corresponding persistence diagram}
\end{figure*}

In this appendix, we briefly explain the procedure to plot a persistence diagram from a network configuration \cite{gameiro2020, mischaikow2013, perseus}.
First, let us consider a force network configuration as in Fig. \ref{fig:11a}, where the numbers represent the force magnitude and the colors represent each connected component.
Now, let us filter the force chains by increasing threshold $\theta_f$, where a link in a network appears when the magnitude is greater than or equal to $\theta_f$.
Once a connected component appears during the filtration, we start to record its appearance in the barcode (Fig. \ref{fig:11b}).
Note that when $\theta_f = 3$, first component (brown) appear, followed by the second component (blue) at $\theta_f = 5$.
These are the birth $\theta_f$ for each connected component.
As $\theta_f$ increases, more chains appear and the connected components grow.
At $\theta_f = 25$, two connected components (blue and green) merge with each other.
When merging of the connected component takes place, we adopt a rule such that a component that is born later in the filtration (which has higher birth $\theta_f$) dies.
In other words, at $\theta_f=25$, component green (birth $\theta_f=15$) dies since it merges with the component blue (birth $\theta_f=5$).
Then, at $\theta_f = 30$, blue component merges with the brown component.
Since $\theta_f = 30$ is the maximum value of the filtration, component brown will never die.
Thus, we consider that it has infinite persistence.
In addition, component black also never die since it does not merge with any other components until the end of the filtration.
Finally, we plot the death and birth $\theta_f$ of each connected component in the persistence diagram (Fig. \ref{fig:11c}), where we assign death $\theta_f = -1$ for the connected components with infinite persistence.
From this example, we demonstrate that persistent homology emphasizes more on the structure of each component rather than its total magnitude since: (i) Green component has higher total magnitude than the blue component, but the blue component has bigger life span (death $\theta_f - $ birth $\theta f$) and (ii) we ignore single link (black component) that is not merged with another component. 
\bibliography{references}
\end{document}